\documentclass[aps,prd,10pt,twocolumn,superscriptaddress,floatfix,nofootinbib,amsmath,amssymb,altaffilletter,preprintnumbers,eprint]{revtex4-2}

\usepackage[normalem]{ulem}
\usepackage{natbib}
\usepackage[english]{babel}
\usepackage[utf8]{inputenc}

\usepackage{enumitem}
\usepackage{tensor}
\usepackage{braket}
\usepackage{footnote}
\usepackage{tablefootnote}
\usepackage{amsmath,amssymb,amsfonts,mathrsfs}
\usepackage[subnum]{cases}
\usepackage{float}
\usepackage{booktabs}
\usepackage{mathtools}
\usepackage{bm, physics}
\usepackage{graphicx}   
\usepackage[caption=false]{subfig}
\usepackage{latexsym}
\usepackage[acronym]{glossaries}
\usepackage{physics}
\usepackage{siunitx}
\usepackage{comment}
\usepackage{fontawesome}
\usepackage{multirow}
\usepackage{dcolumn}
\usepackage[svgnames]{xcolor}
\usepackage[final]{hyperref}

\hypersetup{%
    colorlinks=true,
    citecolor=Blue,
    linkcolor=Blue,
    urlcolor=Blue
    }%

\usepackage{academicons}
\definecolor{orcidlogocol}{HTML}{A6CE39}
\newcommand{\orcid}[1]{\href{https://orcid.org/#1}{\textcolor[HTML]{A6CE39}{\aiOrcid}}}

\begin{document}

\title{Quadratic gravity corrections to scalar QNMs of rapidly rotating black holes}

\author{Stef~J.~B.~Husken}
\email{stef.husken@student.kuleuven.be}
\affiliation{Institute for Theoretical Physics, KU Leuven,
Celestijnenlaan 200D, 3001 Leuven, Belgium}
\affiliation{Leuven Gravity Institute, KU Leuven,
Celestijnenlaan 200D, 3001 Leuven, Belgium}

\author{Tom~van~der~Steen}
\email{tom.vandersteen@kuleuven.be}
\affiliation{Institute for Theoretical Physics, KU Leuven,
Celestijnenlaan 200D, 3001 Leuven, Belgium}
\affiliation{Leuven Gravity Institute, KU Leuven,
Celestijnenlaan 200D, 3001 Leuven, Belgium}

\author{Simon~Maenaut}
\email{simon.maenaut@nbi.ku.dk}
\affiliation{Center of Gravity, Niels Bohr Institute,
Blegdamsvej 17, 2100 Copenhagen, Denmark}

\author{Kelvin~Ka-Ho~Lam}
\affiliation{Illinois Center for Advanced Studies of the Universe \& Department of Physics,
University of Illinois Urbana-Champaign, Urbana, Illinois 61801, USA}

\author{Maxim~D.~Jockwer}
\affiliation{Institute for Theoretical Physics, KU Leuven,
Celestijnenlaan 200D, 3001 Leuven, Belgium}

\author{Adrian~Ka-Wai~Chung}
\affiliation{DAMTP, Centre for Mathematical Sciences, University of Cambridge,
Wilberforce Road, Cambridge CB3 0WA, United Kingdom}

\author{Thomas~Hertog}
\affiliation{Institute for Theoretical Physics, KU Leuven,
Celestijnenlaan 200D, 3001 Leuven, Belgium}
\affiliation{Leuven Gravity Institute, KU Leuven,
Celestijnenlaan 200D, 3001 Leuven, Belgium}

\author{Tjonnie~G.~F.~Li}
\affiliation{Institute for Theoretical Physics, KU Leuven,
Celestijnenlaan 200D, 3001 Leuven, Belgium}
\affiliation{Leuven Gravity Institute, KU Leuven,
Celestijnenlaan 200D, 3001 Leuven, Belgium}
\affiliation{STADIUS Center for Dynamical Systems, Signal Processing and Data Analytics,
KU Leuven, Kasteelpark Arenberg 10, 3001 Leuven, Belgium}

\author{Nicol\'as Yunes}
\affiliation{Illinois Center for Advanced Studies of the Universe \& Department of Physics,
University of Illinois Urbana-Champaign, Urbana, Illinois 61801, USA}

\begin{abstract}
    In an effective-field-theory framework for gravity, black-hole quasinormal mode spectra acquire corrections in quadratic-curvature, scalar-tensor extensions of general relativity. Previous calculations of such corrections were limited to moderate spins, since the corresponding background solutions relied on expansions in the spin parameter. Using recently constructed numerical black-hole solutions valid for large spin, we compute the leading-order deviations from general relativity in the scalar quasinormal mode spectrum of rotating black holes in scalar Gauss-Bonnet and dynamical Chern-Simons gravity. We solve the resulting perturbation equations with pseudo-spectral collocation methods, allowing us to determine the quasinormal-mode corrections for dimensionless spins up to $a/M=0.99$, with accuracy better than $\lesssim 10^{-3}$ for the $l=m=0$ mode and $\lesssim 10^{-6}$ for higher multipoles. For spins $a/M>0.9$, the corrections to certain modes can increase by orders of magnitude.
\end{abstract}

\maketitle

\section{Introduction}

Black holes (BHs) are among the most compact objects in the Universe. In general relativity (GR), stationary BHs are fully characterized by three parameters: mass, spin, and electric charge \cite{no_hair,no_hair2,no_hair3}. However, astrophysical BHs are expected to have negligible electric charge, and are therefore effectively only described by their mass and spin \cite{zajacek2019electricchargeblackholes},  which makes them an ideal theoretical and observational laboratory to probe gravity and a unique testing ground for strong-field effects and physics beyond GR.

In the era of gravitational-wave (GW) astronomy, observations by the LIGO-Virgo-KAGRA collaboration have provided a growing catalogue of binary BH mergers \cite{LVKcollaborationGWTC40UpdatingGravitationalWave2025}. These observations provide a direct empirical window into the strong-field regime of gravity \cite{LVKcollaborationBlackHoleSpectroscopy2026, LVKcollaborationGW250114TestingHawkings2025, LVKcollaborationTestsGeneralRelativity2025,LVKcollaborationTestsGeneralRelativity2026a,LVKcollaborationTestsGeneralRelativity2026b,LVKcollaborationTestsGeneralRelativity2026c}. The final stage of the merger is particularly interesting for testing GR. During this phase, known as the ringdown, the remnant BH settles into its final stationary state and the GW signal carries detailed information about the geometry of the remnant BH \cite{yunesGravitationalwaveTestsGeneral2025, carulloEmpiricalTestsBlack2018, bertiExtremeGravityTests2018}.

The late-time ringdown stage can be modelled using BH perturbation theory, during which the remnant settles down to its final, stationary state through the emission of GWs. The latter can be represented as quasinormal modes (QNMs), exponentially damped sinusoids determined by the characteristic frequencies and damping times of the final BH state \cite{bertiBlackHoleSpectroscopy2025}. The QNM spectrum is completely determined by the background spacetime, making these modes key observables for probing the properties of BHs. From an observational perspective, measuring QNMs enables tests of GR through BH spectroscopy \cite{dreyerBlackholeSpectroscopyTesting2004, bertiGravitationalwaveSpectroscopyMassive2006, bertiQuasinormalModesBlackholes2009, franchiniTestingGeneralRelativity2024, bertiBlackHoleSpectroscopy2025}. From a theoretical perspective, QNMs also play a fundamental role in determining the linear stability of BH solutions, both within GR and in alternative theories of gravity \cite{vishveshwaraStabilitySchwarzschildMetric1970, pressPerturbationsRotatingBlack1973}.

To study theories beyond GR in a systematic fashion, we adopt an effective field theory (EFT) approach to gravity and extend the action of GR with all possible terms compatible with the theory's symmetries, organized in a derivative expansion \cite{donoghueGeneralRelativityEffective1994, endlichEffectiveFormalismTesting2017}. Local Lorentz invariance and diffeomorphism invariance demand that corrections introduced to the action be scalar quantities, and if one restricts attention to modifications that depend only on the metric, these scalars must be constructed from the curvature tensor. Such EFT extensions of GR are sometimes referred to as higher-derivative gravity \cite{stelleClassicalGravityHigher1978,stelleRenormalizationHigherderivativeQuantum1977,canoLeadingHigherderivativeCorrections2019}. From a Wilsonian viewpoint, such higher-derivative corrections can arise from integrating out massive degrees of freedom, which can then couple to lighter or massless degrees of freedom if present in the theory \cite{alexanderCosmologyGraviAxions2025}. More generally, however, related operators can also emerge through anomaly-cancelling mechanisms in the standard model and its extensions \cite{alexanderChernSimonsModified2009}.

In this article, we focus on two theories: scalar Gauss-Bonnet (sGB) gravity and dynamical Chern-Simons (dCS) gravity. These are scalar-tensor theories, in which a scalar field couples to a quadratic curvature invariant (see Sec.~\ref{sec:sGB_dCS}), in contrast to pure higher-curvature theories that do not introduce additional scalar degrees of freedom \cite{endlichEffectiveFormalismTesting2017, cardosoBlackHolesEffective2018}. Both theories are well-motivated: parity-even sGB gravity appears in the low-curvature limit of certain string theories \cite{maedaBlackHoleSolutions2009, mouraHigherDerivativeCorrectedBlack2007}, while parity-odd dCS gravity arises in the low-curvature limit of heterotic string theory \cite{bergshoeffTendimensionalMaxwellEinsteinSupergravity1982, greenAnomalyCancellationsSupersymmetric1984}, and is also connected to gravitational anomalies in particle physics \cite{alexanderChernSimonsModified2009}. 

The calculation of corrections to the QNM spectrum of BHs in higher-curvature theories was considered for non-rotating BHs in \cite{cardosoPerturbationsSchwarzschildBlakcholes2009, molinaGravitationalSignatureSchwarzschild2010, blazquezsalcedoPerturbedBlackHoles2016, blazquezsalcedoQuasinormalmodesEinsteingaussbonnetdilatonBlackholes2017, tattersallQuasinormalmodesBlackholesHorndeski2018, cardosoBlackHolesEffective2018, cardosoParametrizedBlackholeQuasinormal2019, mcmanusParametrizedBlackholeQuasinormal2019, konoplyaQuasinormalModesStability2020, derhamBlackholeGravitationalWaves2020, silvaQuasinormalModesExcitation2024} and at linear or quadratic order in the spin in \cite{pieriniQuasinormalmodesRotatingBlackholes2021, srivastavaAnalyticalComputationQuasinormalmodes2021, canoGravitationalRingingRotating2022, pratikQuasinormalModesSlowlyrotating2022, pieriniQuasinormalmodesRotatingBlackholes2022}. To compute QNMs of rapidly spinning black holes in modified gravity the difficulty comes from the fact that the background geometry is modified relative to Kerr and the perturbation equations themselves are also modified. Broadly speaking, two classes of methods have been developed to address this problem. In field-perturbation approaches, one perturbs the modified black-hole background with time-dependent fields and solves the resulting coupled partial differential equation (PDE) eigenvalue problem numerically, often with spectral or pseudo-spectral methods \cite{hussainApproachComputingSpectral2022,chungSpectralMethodGravitational2023,chungSpectralMethodMetric2024,blazquez-salcedoQuasinormalModesKerr2024,chungRingingOutGeneral2024,chungQuasinormalModeFrequencies2024,chungQuasinormalModeFrequencies2025}. In curvature-perturbation approaches, one instead expands suitable Newman-Penrose scalars and derives modified Teukolsky-type equations within an EFT treatment \cite{liPerturbationsSpinningBlack2023,canoUniversalTeukolskyEquations2023,canoParametrizedQuasinormalMode2024,waglePerturbationsSpinningBlack2024,canoHigherderivativeCorrectionsKerr2024,boyceKerrBlackHole2026}. For this paper we draw inspiration from both methods, restricting ourselves to scalar modes and using pseudo-spectral methods. %follow the first route, using pseudo-spectral methods.

Existing calculations of QNMs in sGB and dCS are restricted to moderately spinning BHs \cite{chungQuasinormalModeFrequencies2024, chungQuasinormalModeFrequencies2025}, since the corresponding background solutions were obtained through expansions in the spin \cite{yunesDynamicalChernSimonsModified2009, yagiSlowlyRotatingBlack2012, ayzenbergSlowlyRotatingBlack2014, maselliRotatingBlackHoles2015, canoLeadingHigherderivativeCorrections2019, CanoAccuracySlowrotationApproximation2024, canoParametrizedQuasinormalMode2024}. However, astrophysical merger remnants are expected to have high spins \cite{loustoRemnantMassesSpins2010, LVKcollaborationGWTC40UpdatingGravitationalWave2025}, where these expansions start to break down. Moreover, recent studies suggest that near-extremal BHs may amplify deviations from GR, making them particularly promising targets for observational tests \cite{horowitzamplification,canoTeukolskyEquationNearextremal2024,canoAmplificationNewPhysics2025}. It is therefore essential to develop methods for computing QNM corrections in modified gravity theories that remain valid at high spin.

Recent progress in constructing BH solutions in higher-derivative gravity for high spins, using spectral methods, now enables the extension of the corrections of the QNM spectrum to large spins \cite{lamSpinningBlackHoles2026, lamAnalyticAccurateApproximate2026, fernandesLeadingEffectiveField2025}. In this work, we focus on scalar QNMs and compute corrections to their spectrum using these new numerical backgrounds. Employing pseudo-spectral techniques, we determine the leading-order corrections to scalar QNMs for dimensionless spins up to $a/M=0.99$. Our approach applies to any modified background perturbatively close to Kerr (see \cite{vandersteenRingingRapidlyRotating2026}) and demonstrates the potential of pseudo-spectral methods for computing QNMs in theories beyond GR.

For moderate spins of $a/M\lesssim0.7$, results obtained using spin-expanded backgrounds agree with the more accurate numerical backgrounds. For higher spins, however, the spin expansions fail to be accurate enough, demonstrating the necessity of a fully numerical treatment. Notably, similar to \cite{canoEikonalQuasinormalModes2025, canoAmplificationNewPhysics2025}, we observe that the corrections to certain modes grow strongly for large spin, in the vicinity of the zero-damped mode to damped mode transition, signalling a significant amplification of beyond-GR effects.

This article is organized as follows. In Sec.~\ref{sec:method} the theoretical framework of modified scalar QNMs will be provided, along with an outline of the numerical methods. In Sec.~\ref{sec:results} we present the numerical results and validate them against values for low spins found in the literature. To conclude, Sec.~\ref{sec:disc} relates our findings to similar results, obtained in the Eikonal limit in \cite{canoEikonalQuasinormalModes2025, canoAmplificationNewPhysics2025}.

\section{Scalar QNMs in scalar Gauss-Bonnet and dynamical Chern-Simons}\label{sec:method}
To study scalar perturbations of axisymmetric spacetimes in quadratic scalar-tensor theories of gravity, we first briefly review scalar Gauss-Bonnet and dynamical Chern-Simons gravity and the corrections they introduce to the Kerr metric. We then proceed in deriving the perturbative equations that a scalar perturbation satisfies on such a modified Kerr background. Finally, we comment on the pseudo-spectral methods used to numerically solve the resulting equations.

\subsection{Modified Kerr BHs in sGB and dCS}\label{sec:sGB_dCS}
Here, we focus on the EFT extensions which are quadratic in the curvature.  In this paper, we look at the specific cases of sGB and dCS gravity. We denote the length scale where these corrections appear as $\ell$. The Lagrangian density for these theories is described by \cite{fundimplications}:
\begin{equation}
    16\pi\mathcal{L}=R-\frac{1}{2}\nabla_\mu\varphi\nabla^\mu\varphi -V(\varphi)+\alpha\: f(\varphi)\mathcal{Q},
\end{equation}
where $V$ is a potential for the scalar field, $\alpha$ a coupling constant of dimension $\ell^2$, $f$ a coupling function and $\mathcal{Q}$ is a quadratic topological invariant. The scalar field $\varphi$ has dimension $\ell$. In the case of sGB, $\varphi$ couples to the parity-even Gauss-Bonnet invariant $\mathcal{Q}_\mathrm{sGB}$, and for dCS it couples to the parity-odd Pontryagin invariant $\mathcal{Q}_\mathrm{dCS}$, which are given by:
\begin{align}
    \mathcal{Q}_\mathrm{sGB} &= R\indices{_{\mu\nu\rho\sigma}}R\indices{^{\mu\nu\rho\sigma}} - 4 R_{\mu\nu} R^{\mu\nu} + R^2,\\
    \mathcal{Q}_\mathrm{dCS} &= \tilde{R}\indices{_{\mu\nu\rho\sigma}}R\indices{^{\mu\nu\rho\sigma}},
\end{align}
where we have the dual Riemann tensor $\tilde{R}\indices{_{\mu\nu\rho\sigma}}=\frac{1}{2}\epsilon\indices{_{\mu\nu\alpha\beta}}R\indices{_{\rho\sigma}^{\alpha\beta}}$. The freedom in choosing the potential $V$ and coupling $f$ still allows for a family of theories. In line with \cite{canoLeadingHigherderivativeCorrections2019, lamAnalyticAccurateApproximate2026, lamSpinningBlackHoles2026}, we restrict to massless, shift-symmetric scalar fields, which corresponds to the choice $V=0$ and $f(\varphi)=\varphi$. This can be interpreted as the small-coupling limit $\alpha/L^2\ll1$ of a general coupling to the topological invariant $\mathcal{Q}$, where $L$ is the characteristic curvature scale of the system \cite{chungQuasinormalModeFrequencies2024}.

We now consider axisymmetric BHs in these quadratic theories with ADM mass $M$ and spin $a$. The characteristic length scale of the spacetime in this case is given by the mass $M$. Hence, we can introduce a dimensionless coupling parameter $\lambda = \alpha^2/M^4$. We work perturbatively in $\lambda$, and expand the metric as:
\begin{equation}
    g_{\mu\nu}=g_{\mu\nu}^{(0)}+\lambda g_{\mu\nu}^{(1)}+\mathcal{O}(\lambda^2).
\end{equation}
The labels $(k)$ indicate the order $k$ in the $\lambda$-expansion, so $g_{\mu\nu}^{(0)}$ is the Kerr metric in GR, while $g_{\mu\nu}^{(1)}$ contains the EFT corrections. In Boyer-Lindquist coordinates $x^\mu = (t,r,y,\phi)$ with $y=\cos\theta$, this modified Kerr metric takes the form \cite{canoLeadingHigherderivativeCorrections2019}:
\begin{equation}\label{eq:background}
\begin{split}
	g_{\mu\nu}dx^\mu dx^\nu =&
    -\left(1-\frac{2Mr}{\Sigma}-\lambda H_1\right) dt^2 \\
	& - (1+\lambda H_2)\frac{4 M ar (1-y^2)}{\Sigma} dt d\phi\\
	&+(1+\lambda H_3)\Sigma\left(\frac{dr^2}{\Delta}+\frac{dy^2}{1-y^2}\right)\\
	&+(1+\lambda H_4)\left(r^2+a^2+\frac{2a^2Mr(1-y^2)}{\Sigma}\right)\\
    &\times(1-y^2)d\phi^2,
\end{split}
\end{equation}
where we have introduced $\Delta = r^2-2Mr+a^2$ and $\Sigma = r^2 + a^2y^2$. The higher-derivative corrections $H_i(r,y)$ with $i\in\{1,2,3,4\}$ only depend on the radial and polar coordinate. This parametrisation for the corrections preserves several properties of a Kerr BH: the horizons are still located at the roots of $\Delta$, i.e. $r_\pm = (M\pm\sqrt{M^2-a^2})$, and $M$ and $a$ are still the ADM mass and spin, respectively, upon imposing appropriate asymptotic conditions on $H_i$ (see \cite{canoLeadingHigherderivativeCorrections2019}).

\subsection{Klein-Gordon Equation on Modified Kerr Backgrounds}

Consider a massless scalar test field $\psi$ propagating on this modified Kerr background.\footnote{A perturbation of the background scalar field $\varphi$ would couple to the invariant $\mathcal{Q}$, and consequently to metric perturbations.} This test field satisfies the Klein-Gordon equation
\begin{equation}\label{eq:klein-gordon}
    \Box\psi(t,r,\phi,y) = 0.
\end{equation}
Since the background metric is stationary and axisymmetric, it admits the Killing vectors $\partial_t$ and $\partial_\phi$. As a consequence, the Klein-Gordon equation can readily be separated as \cite{canoRingingRotatingBlack2020}:
\begin{equation}\label{eq:separation}
    \Box \psi = \int_{-\infty}^\infty d\omega \sum_{m=-\infty}^\infty e^{i(m\varphi-\omega t)}\mathcal{D}_m(\omega)[\psi_{m,\omega}]=0,
\end{equation}
where $m\in\mathbb{Z}$ due to the periodic boundary conditions on $\phi$. Here, we have introduced a new differential operator $\mathcal{D}_m(\omega)$, which contains derivatives up to second order in $r$ and $y$. Now, for a given $m$ and $\omega$, each component $\psi_{m,\omega}(r,y)$ should satisfy the Laplace equation
\begin{equation}\label{eq:Laplace}
    \mathcal{D}_{m}(\omega)[\psi_{m,\omega}] = 0,
\end{equation}
in order for the scalar field to obey the Klein-Gordon Eq.~\eqref{eq:klein-gordon}. Since our background is perturbative in $\lambda$, the Laplace operator $\mathcal{D}_m(\omega)$ admits the same expansion and splits up in a GR operator plus corrections:
\begin{equation}\label{eq:D_expanded}
    \mathcal{D}=\mathcal{D}^{(0)}+\lambda\mathcal{D}^{(1)}+\mathcal{O}(\lambda^2),
\end{equation}
where, from this point on, we will omit the indicators for $m$ and $\omega$ to prevent notational clutter. 

Since we are already working perturbatively in $\lambda$, it is natural to assume the corrected QNMs are close to their GR counterparts. This is also supported by current observations of gravitational QNMs, which are still consistent with their GR values \cite{LVKcollaborationTestsGeneralRelativity2025, LVKcollaborationBlackHoleSpectroscopy2026, LVKcollaborationTestsGeneralRelativity2026b, LVKcollaborationTestsGeneralRelativity2021, carulloEnhancingModifiedGravity2021, silvaBlackholeRingdownProbe2023, julieInspiralmergerringdownWaveforms2025, liuRobustImprovedConstraints2025, maenautRingdownAnalysisRotating2026, chungProbingQuadraticGravity2025a}. Hence, we also expand
\begin{align}
	\psi(r,y)&=\psi^{(0)}(r,y)+\lambda \psi^{(1)}(r,y)+\mathcal{O}(\lambda^2)
    \label{eq:psi_expanded}\\
    \omega&=\omega^{(0)}+\lambda \omega^{(1)}+\mathcal{O}(\lambda^2)
    \label{eq:om_expanded}
\end{align}
Now we insert the expansions Eq.~\eqref{eq:D_expanded}--\eqref{eq:om_expanded} into Eq.~\eqref{eq:separation}. Since the Killing vectors $\partial_t$ and $\partial_\phi$ are still preserved as symmetries of the corrected background Eq.~\eqref{eq:background}, we can still separate the modes $e^{i(m\varphi-\omega t)}$, before performing the expansion in $\omega$. After separation, we expand Eq.~\eqref{eq:Laplace}, and upon ordering the powers of $\lambda$ we obtain  the system of equations
\begin{numcases}{}\label{eq:D_eqs}
    \mathcal{D}^{(0)}[\psi^{(0)}] & $= 0$, 
    \label{eq:Eq_O0}\\
    \mathcal{D}^{(0)}[\psi^{(1)}] & $=-\omega^{(1)}\delta\mathcal{D}^{(0)}[\psi^{(0)}]-\mathcal{D}^{(1)}[\psi^{(0)}]$,
    \label{eq:Eq_O1}
\end{numcases}
where all the operators depend on the GR-value $\omega^{(0)}$ and where we have introduced a new operator $\delta\mathcal{D}^{(0)}$ defined by 
\begin{equation}
    \delta\mathcal{D}^{(0)}[f]\equiv \left. \left(\frac{\partial}{\partial\omega}\mathcal{D}^{(0)}\right) \right\lvert_{\omega=\omega^{(0)}}[f],
\end{equation}
for an arbitrary field $f$. Eq.~\eqref{eq:Eq_O0} is the equation for scalar perturbations on a Kerr background, for which the solutions are well-understood due to its coordinate separability \cite{teukolskyRotatingBlackHoles1972, teukolskyPerturbationsRotatingBlack1973, leaverAnalyticRepresentationQuasinormal1985}. By first solving the GR quantities $\omega^{(0)}$ and $\psi^{(0)}$, one can then solve for the corrections using Eq.~\eqref{eq:Eq_O1}, where the GR solutions enter as source terms. This equation is non-separable and thus much harder to solve \cite{canoRingingRotatingBlack2020, miguelEFTCorrectionsScalar2024}. 

In their current form, Eqs.~\eqref{eq:Eq_O0}--\eqref{eq:Eq_O1} exhibit divergent behaviour at the boundaries of the domain and do not enforce the proper boundary conditions (BCs) to be a QNM: being purely ingoing at the horizon and outgoing at infinity. This is equivalent to imposing that no radiation emerges from the horizon and no radiation enters from spatial infinity. To recast these equations in a form that is numerically solvable, we first compactify the domain and then factor out the right behaviour at the boundaries. 

We compactify the radial coordinate\footnote{Note that the sources for the spectral backgrounds use a slightly different coordinate $x \equiv 2 \, (\flatfrac{r_+}{r})-1$ where the BH event horizon corresponds to $x=1$ and spatial infinity to $x=-1$ \cite{lamSpinningBlackHoles2026, lamAnalyticAccurateApproximate2026}.} as follows \cite{leaverAnalyticRepresentationQuasinormal1985, cookGravitationalPerturbationsKerr2014}: %, miguelEFTCorrectionsScalar2024
\begin{equation}
    z \equiv \frac{r-r_+}{r-r_-}.
\end{equation}
This coordinate maps the event horizon to $z=0$ and spatial infinity to $z=1$, compactifying the BH exterior $r\in[r_+,\infty)$ to $z\in[0,1]$. 

To enforce the BCs for a QNM, we explicitly factor out a regularizing function $A_{m,\omega}$ that captures the correct QNM behaviour at the event horizon and infinity, i.e. we write
\begin{equation}
    \psi(z,y) \equiv A_{m,\omega}(z,y)\tilde{\psi}(z,y),
\end{equation}
where $\tilde{\psi}(z,y)$ is a regular function that is smooth on the domain $z\in[0,1]$ and $y\in[-1,1]$. 
Note that the modified Kerr background Eq.~\eqref{eq:background} has a different asymptotic structure and different behaviour at the horizon, compared to the Kerr solution in GR \cite{canoLeadingHigherderivativeCorrections2019}. 
Hence, we also expand $A_{m,\omega}$  and $\tilde{\psi}(z,y)$ up to linear order to accommodate for changes in the BCs, 
\begin{equation}
\begin{split}
A_{m,\omega} & = A^{(0)}_{m,\omega} + \lambda A^{(1)}_{ m,\omega}, \\
\tilde{\psi}_{m,\omega}(z,y) &= \tilde{\psi}^{(0)}_{m,\omega}(z,y) + \lambda \tilde{\psi}^{(1)}_{m,\omega}(z,y), 
\end{split}
\end{equation}
so that we can write
\begin{equation}
    \psi^{(0)}\equiv A^{(0)}\tilde{\psi}^{(0)}, \quad \psi^{(1)}\equiv A^{(0)}\tilde{\psi}^{(1)} + A^{(1)}\tilde{\psi}^{(0)}.
\end{equation}
where the subscripts $m$ and $\omega$ will be suppressed from now on. Inserting this regularization in Eqs.~\eqref{eq:Eq_O0}--\eqref{eq:Eq_O1}, we obtain the regular system of equations:
\begin{widetext}
    \begin{numcases}{}
        \tilde{\mathcal{D}}^{(0)}(\omega^{(0)})[\tilde{\psi}^{(0)}] & $=0$, 
        \label{eq:Eq_O0_reg}\\
        \tilde{\mathcal{D}}^{(0)}(\omega^{(0)})[\tilde{\psi}^{(1)}] & $= -\tilde{\mathcal{D}}^{(0)}(\omega^{(0)})[ B^{(1)}\tilde{\psi}^{(0)}] - \omega^{(1)}\tilde{\delta\mathcal{D}}^{(0)}[\tilde{\psi}^{(0)}]-\tilde{\mathcal{D}}^{(1)}(\omega^{(0)})[\tilde{\psi}^{(0)}]$,
        \label{eq:Eq_O1_reg}
    \end{numcases}
\end{widetext}
where a tilde indicates that an operator has been regularized with respect to $A^{(0)}$. For instance, one has
\begin{equation}\label{eq:D0_reg}
    \tilde{\mathcal{D}}^{(0)}[\tilde{\psi}]\equiv\frac{1}{A^{(0)}}\mathcal{D}^{(0)}[A^{(0)}\tilde{\psi}],
\end{equation}
Furthermore, we have also introduced the ratio
\begin{equation}
    B^{(1)}(z,y)\equiv \frac{A^{(1)}(z,y)}{A^{(0)}(z,y)}.
\end{equation}
The GR regulator for the Kerr operator is a well-known expression, that reflects the operators separability \cite{leaverAnalyticRepresentationQuasinormal1985, cookGravitationalPerturbationsKerr2014, miguelEFTCorrectionsScalar2024}: it splits in the radial component
\begin{equation}\label{eq:A0_radial}
    A_r^{(0)}(r)\equiv e^{i\omega r}(r-r_+)^{-i\sigma_+}(r-r_-)^{-1 + 2iM\omega +i\sigma_+},
\end{equation}
where $\sigma_+=(2M\omega r_+-am)/(r_+-r_-)$, which is related to the surface gravity and horizon angular momentum (see \cite{canoParametrizedQuasinormalMode2024, chungQuasinormalModeFrequencies2024}) and the angular component
\begin{equation}\label{eq:A0_angular}
    A_y^{(0)}(y)\equiv(1-y)^{|m|/2}(1+y)^{|m|/2}.
\end{equation}
This yields an operator $\tilde{\mathcal{D}}^{(0)}$ that has smooth coefficients. Hence, Eq.~\eqref{eq:Eq_O0_reg} is regular and solutions to it satisfy the BCs for a QNM. Eq.~\eqref{eq:Eq_O1_reg}, on the other hand, is not readily regularized; the operators $\tilde{\delta\mathcal{D}}^{(0)}$ and $\tilde{\mathcal{D}}^{(1)}$ have coefficients that diverge on the boundaries. Imposing the boundary conditions is equivalent to demanding the cancellation of divergent terms, which fixes $B^{(1)}$ (see \cite{vandersteenRingingRapidlyRotating2026} for the detailed description of finding the correct ratio).

With this procedure we have obtained a regular system of partial differential equations (PDEs) in Eqs.~\eqref{eq:Eq_O0_reg}--\eqref{eq:Eq_O1_reg} that satisfies the correct BCs. Any smooth solution $\tilde{\psi}$ that is non-zero at the boundaries is guaranteed to be a scalar QNM of the modified Kerr BH given in Eq.~\eqref{eq:background}. In its current formulation, the problem is well-suited to be solved numerically with pseudo-spectral techniques.

\subsection{Pseudo-spectral methods}\label{sec:pseudospectral}

In Eqs.~\eqref{eq:Eq_O0_reg}--\eqref{eq:Eq_O1_reg}, we have obtained a system of PDEs with smooth coefficients, that defines an eigenvalue problem for $(\omega^{(0)},\omega^{(1)})$. Since the coefficients depend non-linearly on the eigenvalue $\omega_{(0)}$, we employ a pseudo-spectral collocation method to solve this problem. Using this method, we express the PDE on a grid for a given set of interpolation functions, which reduces differentiation operators to discrete matrices \cite{grandclementSpectralMethodsNumerical2009, diasNumericalMethodsFinding2016, canutoSpectralMethodsFundamentals2006, boydChebyshevFourierSpectral2001}. Consequently, the continuous differential equation is recast as a discrete linear algebra problem. Our implementation closely follows the framework laid out in \cite{miguelEFTCorrectionsScalar2024}.

We implement our collocation method using Chebyshev polynomials in both the radial and angular sector, which provides exponential convergence \cite{boydChebyshevFourierSpectral2001}. These interpolation functions are expressed on a Gauss–Lobatto grid, which clusters at the boundaries of the domain. This is well-suited for the boundary value problem we are solving. Concretely, we implement the following grid for the radial coordinate:
\begin{equation}
    z_i=\frac{1}{2}[1-\cos(i\pi/N_r)], \quad i=0,\ldots,N_r.
\end{equation}
For the angular coordinate $y$, we use
\begin{equation}
    y_j = \cos(j\pi/N_y), \quad j=0,\ldots,N_y.
\end{equation}
Therefore, we essentially reduce all continuous vectors $\tilde{\psi}$ to a vector $\mathbf{\tilde{\Psi}}_{i,j}\equiv\tilde{\psi}(z_i,y_j)$, which can be flattened to a vector of length $N_r\cdot N_y$. Similarly, the differential operators are tensor products of radial and angular differentiation matrices, which give matrices of size $(N_r\cdot N_y)\times (N_r\cdot N_y)$.

We first solve the zeroth order equation for $(\mathbf{\tilde{\Psi}}^{(0)}, \omega^{(0)})$, which determines the source terms in the first order equation that we solve for $(\mathbf{\tilde{\Psi}}^{(1)}, \omega^{(1)})$. Since at zeroth order in Eq.~\eqref{eq:Eq_O0_reg} the problem is separable, we can reduce it to a separate radial and angular equation that are coupled by the frequency $\omega_{(0)}$ and separation constant $A_{lm}$. Hence, the discrete system of equations is of the form:
\begin{numcases}{}
    \mathbf{L}_r(\omega^{(0)},A_{lm})\cdot\mathbf{R} & $=0$,
    \label{eq:RTE_discrete}\\
    \mathbf{L}_y(\omega^{(0)},A_{lm})\cdot\mathbf{\Theta} & $=0$,
    \label{eq:ATE_discrete}
\end{numcases}
where $\mathbf{R}$ and $\mathbf{\Theta}$ are vectors of length $N_r$ and $N_y$, respectively. The matrices $\mathbf{L}_r$ and $\mathbf{L}_y$ are discretized differential operators of the shape $N_r\times N_r$ and $N_y\times N_y$, respectively. In the Schwarzschild limit, the angular equation is solved analytically and the radial equation reduces to a standard numerical quadratic eigenvalue problem. For finite spin, the coupled system is solved iteratively by incrementally increasing the spin and using a Newton–Raphson procedure applied to the equations for 
$(\mathbf{R}, \mathbf{\Theta},\omega^{(0)},A_{lm})$ (see Eq.~(33) of \cite{vandersteenRingingRapidlyRotating2026}). In practice, we do not need to iterate in spin. Instead, for a given spin we start from tabulated values for $\omega^{(0)}$ (e.g. \cite{bertiExtremeGravityTests2018}) as accurate initial guesses. Now, only a few Newton-Raphson iterations are required to obtain the corresponding eigenvectors to the desired precision. This procedure yields the zeroth order solutions $\omega_{(0)}$ and $\mathbf{\tilde{\Psi}}^{(0)}=\mathbf{R}\otimes\mathbf{\Theta}$. These solutions then serve as input for the computation of higher-order corrections. At this order, we solve the non-separable, non-homogeneous Eq.~\eqref{eq:Eq_O1_reg} on the full grid for $(\mathbf{\tilde{\Psi}}^{(1)}, \omega^{(1)})$, which is equivalent to solving a linear problem:
\begin{equation}
    \mathbf{M}\cdot\mathbf{x}=\mathbf{s}, 
\end{equation}
with
\begin{align}
    \mathbf{M}&=\left(
        \begin{matrix}
           \mathbf{\tilde{D}}^{(0)} & (\boldsymbol{\delta}\mathbf{\tilde{D}}^{(0)}\cdot \mathbf{\tilde{\Psi}}^{(0)})^T\\
           (\mathbf{\tilde{\Psi}}^{(0)})^T & \mathbf{0} 
        \end{matrix}
    \right)\nonumber\\
    \mathbf{x}&=\left(
        \begin{matrix}
            \mathbf{\tilde{\Psi}}^{(1)} & \omega^{(1)}
        \end{matrix}
    \right)^T \nonumber\\
    \mathbf{s}&=\left(
        \begin{matrix}
            -\mathbf{\tilde{D}}^{(1)}\cdot\mathbf{\tilde{\Psi}}^{(0)} & 0
        \end{matrix}
    \right)^T.
\end{align}
The additional row enforces the orthogonality condition $(\mathbf{\tilde{\Psi}}^{(0)})^T\mathbf{\tilde{\Psi}}^{(1)}$=0, which ensures we find a unique solution (see Sec.~III.B of \cite{vandersteenRingingRapidlyRotating2026}). The operators $\mathbf{\tilde{D}}$ are discretized versions of the regular, continuous operators $\tilde{\mathcal{D}}$ from Eq. (\ref{eq:Eq_O1_reg}) with any divergent coefficients removed by the $B^{(1)}$ term. This system can be solved for $\mathbf{x}$ using standard linear solve methods, like Mathematica's \texttt{LinearSolve}, to find the scalar QNM corrections.

\section{Results}\label{sec:results}

We apply the pseudo-spectral solver to the sGB as well as dCS quadratic theories. For both theories, we determine the scalar QNM corrections up to dimensionless spin $a/M=0.99$, corresponding to the spectral backgrounds available from \cite{lamSpinningBlackHoles2026, lamAnalyticAccurateApproximate2026}. In general, we have used the highest resolution background solutions with spectral order 45 in both radial and angular coordinate. A precision of 80 digits has been used. Without loss of generality we set $M=1$ in what follows. For both theories, we have calculated the QNMs for $l\le5$ and $|m|\le l$. We only cover the fundamental $n=0$ modes, but we were also able to track the first overtone $n=1$ for the $l=m=2$ mode. The full list of results will be made public upon publication through a GitHub repository\footnote{\faGithub\hspace{0.2em}\href{https://github.com/StefHusken/scalar-QNMs-higher-derivative-gravity}{scalar QNMs higher-derivative gravity}}.

\subsection{Scalar Gauss-Bonnet}

\begin{figure*}%
    \centering
    \subfloat[sGB]{
    \includegraphics[width=0.48\linewidth]{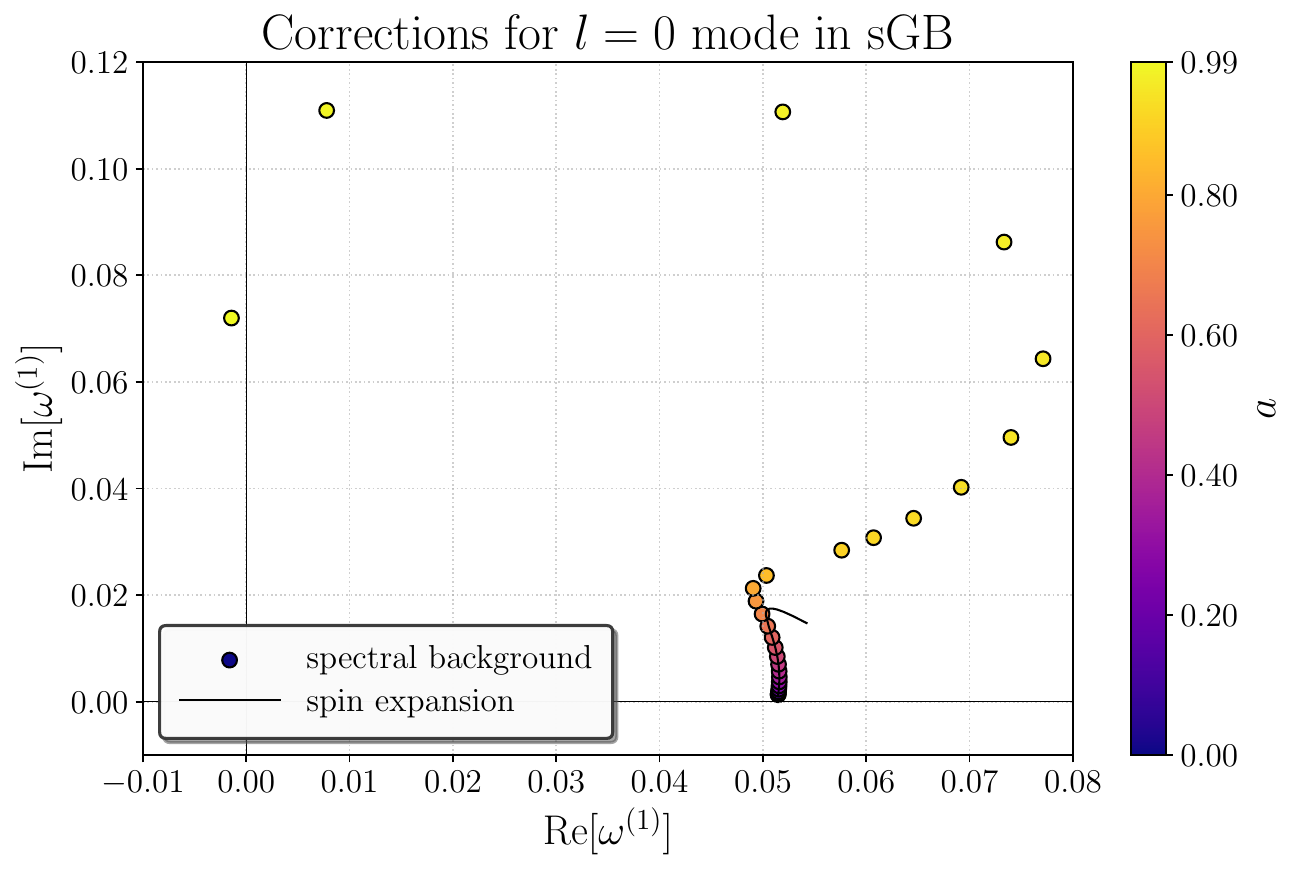}
    \label{fig:l0sGB}
    }%
    \hfill
    \subfloat[dCS]{%
    \includegraphics[width=0.48\linewidth]{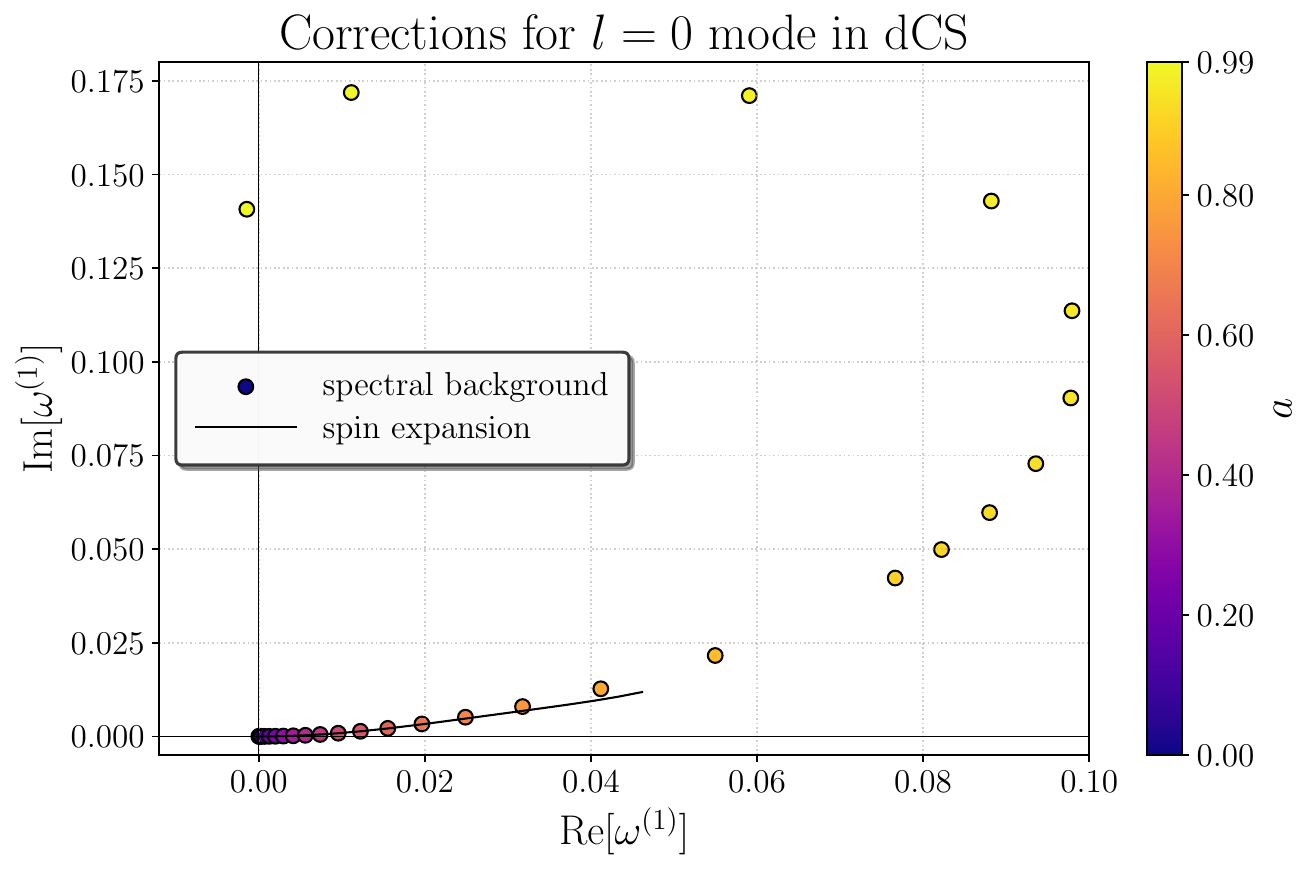}
    \label{fig:l0dCS}
    }%
    \caption{Lowest order corrections $\omega^{(1)}$ to $l=m=0$ scalar QNMs, visualized in the complex plane. Results that were obtained using the spin-expanded background  (lines), shown up until $a=0.8$, are compared to results for the spectral background (circles). The circles have a colour-scale to indicate their corresponding spin, going from $a=0$ until $a=0.99$.}
    \label{fig:l0_comparison}
\end{figure*}

The leading order corrections $\omega^{(1)}$ for the $l=m=0$ mode are displayed in the complex plane in Fig.~\ref{fig:l0sGB}. The spins $a$ are incremented in steps of $\Delta a = 0.05$ for spins up to $0.9$ and in steps of $\Delta a=0.01$ for the higher spins above $a=0.9$ because the changes to QNM frequencies are more rapid for $a$ close to extremality. 
We also include the results obtained by using the background solution in the form of a spin expansion up to order $\mathcal{O}(a^{14})$ from \cite{canoLeadingHigherderivativeCorrections2019}, for comparison. We see that at lower spins $a\lesssim0.7$, the spin expansion is accurate enough to yield corrections consistent with the more accurate spectral background. However for higher spins, the results for the spin expansion start to be inaccurate. The spin expansion even produces the wrong kind of behaviour, as it turns the opposite direction for large spins compared to the more accurate results for the numerical background.

At low spins, we have verified the results using an eigenvalue perturbation (EVP) method \cite{zimmermanQuasinormalModesKerr2015, hussainApproachComputingSpectral2022,canoQuasinormalModesRotating2023}. The EVP method applied to the spin expansion background is in excellent agreement with our pseudo-spectral method applied to the spectral background for spins up to $a=0.6$. 

\begin{figure*}[t]
\centering
\includegraphics[width=\linewidth]{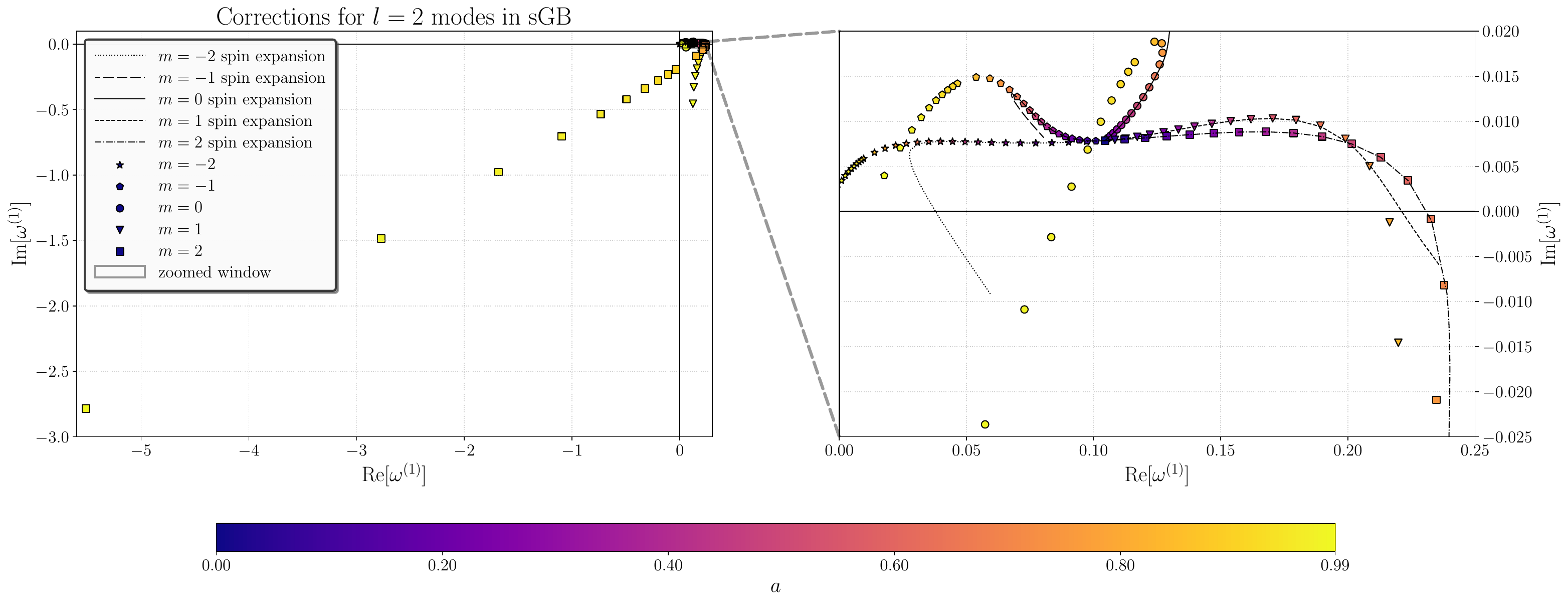}
\caption{Lowest order corrections to $l=2$ scalar QNMs in sGB for different $m$, visualized in the complex plane. A comparison is made between results (up until $a=0.8$) using spin expansion as background spacetime and results for spins up to $a=0.99$, using the spectral background. The lines show spin expansion results. A dotted line represents $m=-2$, a long dashed line $m=-1$, a full line $m=0$, a short dashed line $m=1$, and a dot dash line $m=2$. Markers show the new results using a numerically calculated background. A star represents $m=-2$, a pentagon $m=-1$, a circle $m=0$, a triangle $m=1$, and a square $m=2.$ The colour denotes the spin of the background black hole, going from $a=0$ until $a=0.99.$}
\label{fig:l2sGB}
\end{figure*}

As an example of a mode with non-zero $l$, we display the $l=2$ corrections in Fig.~\ref{fig:l2sGB}. These corrections show some striking behaviour that we also observe for several other $l\neq 0$ modes. Some corrections $\omega^{(1)}_{lm}$ particular $l$ and $m$ values show signs of a divergence as we start to approach the near-extremal regime. The $\ell=m=2$ mode in Fig~\ref{fig:l2sGB}, for example, is orders of magnitude bigger at $a=0.99$, compared to its value at $a=0.9$. This seemingly divergent behaviour also occurs for $l=1$ at $m=1$, for $l=3$ at $m=2,3$, for $l=4$ at $m=3,4$, and for $l=5$ at $m=3,4,5$. This immense growth either signals an amplification of beyond GR effects or a potential breakdown of the leading-order perturbative treatment. We will comment on this in more detail in Sec.~\ref{sec:divergent} and the Discussion in Sec.~\ref{sec:disc}.

To validate our numerical methods and verify that the amplification of certain modes is not a numerical artifact, we confirm the convergence of our results. In Fig.~\ref{fig:l2convergencesGB}, we depict the accuracy of the correction $\omega^{(1)}$ for the fundamental $l=m=2$ mode as a function of the grid size. Here, we display the convergence of the solutions for several spins, where the angular grid is always taken to be $20$ grid points less than the radial grid size $N_r$. It is clear that the solutions do indeed converge exponentially as expected for pseudo-spectral methods. Although, the solutions for higher spins do converge slower than their lower spin counterparts. Results are obtained with relative errors smaller than $10^{-3}$ for all spins and modes, with $l=m=0$ being the most inaccurate solution. For $l>1$ relative errors are at least below $10^{-6}$ for all spins, but in many cases even lower. We note that in general higher multipoles converge faster.

We have also repeated some of the QNM calculations with a lower resolution background. For a background with a spectral order of $40$ (instead of $45$) in both coordinates, we recover shifts in the corrections of orders less than $10^{-16}$ in all tested cases. This indicates that it is the resolution of the QNM solver that is the limiting factor, and not the resolution of the background.

\begin{figure*}[t]
\centering
\includegraphics[width=\linewidth]{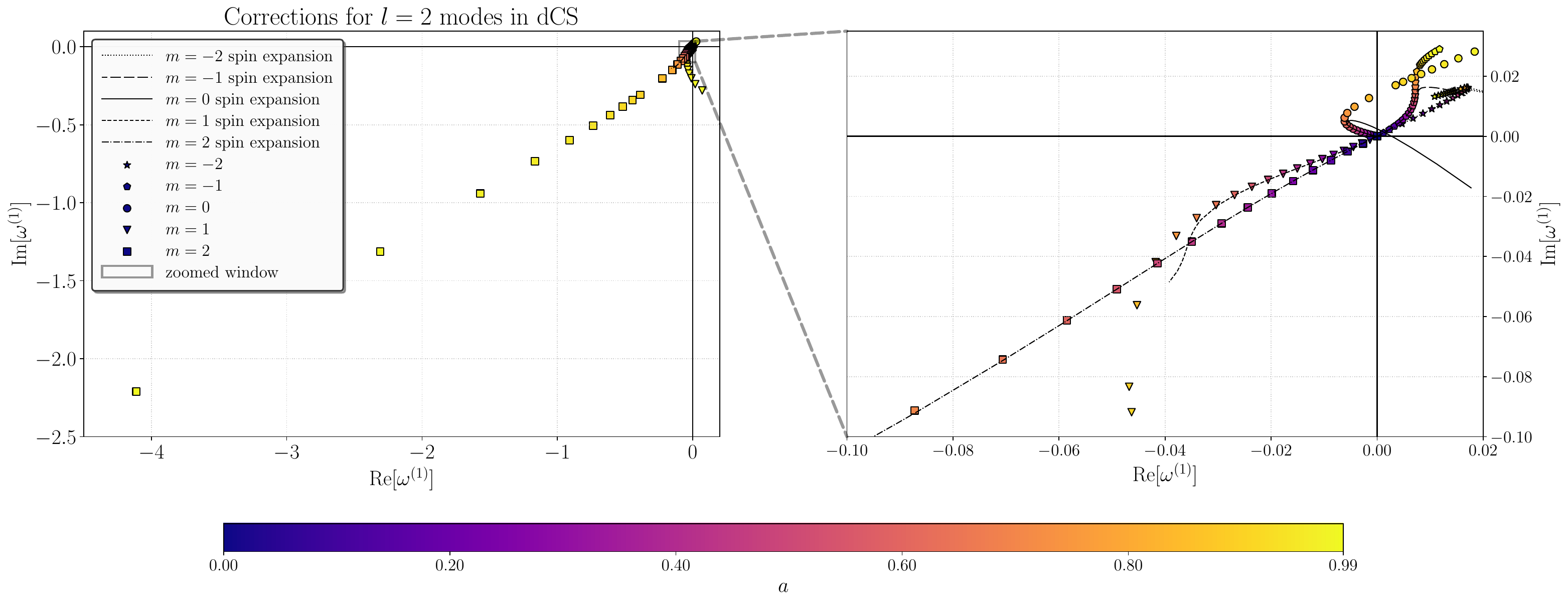}
\caption{Lowest order corrections to $l=2$ scalar QNMs in dCS for different $m$, visualized in the complex plane. A comparison is made between results (up until $a=0.8$) using spin expansion as background spacetime and results for spins up to $a=0.99$, using the spectral background. The lines show spin expansion results. A dotted line represents $m=-2$, a long dashed line $m=-1$, a full line $m=0$, a short dashed line $m=1$, and a dot dash line $m=2$. Markers show the new results using a numerically calculated background. A star represents $m=-2$, a pentagon $m=-1$, a circle $m=0$, a triangle $m=1$, and a square $m=2.$ The colour denotes the spin of the background black hole, going from $a=0$ until $a=0.99.$}
\label{fig:l2dCS}
\end{figure*}

\begin{figure*}%
    \centering
    \subfloat[sGB]{
    \includegraphics[width=0.48\linewidth]{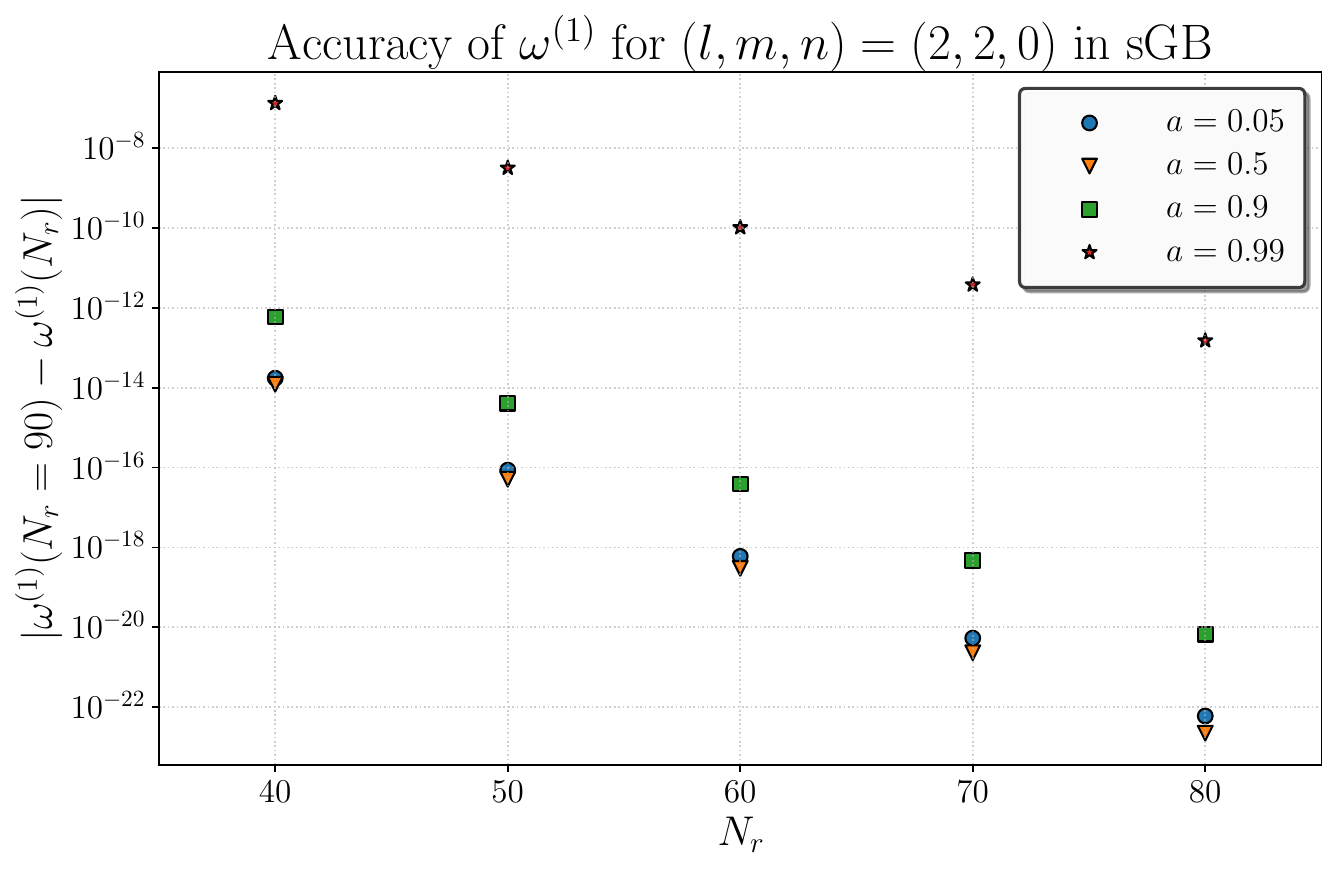}
    \label{fig:l2convergencesGB}
    }%
    \hfill
    \subfloat[dCS]{%
    \includegraphics[width=0.48\linewidth]{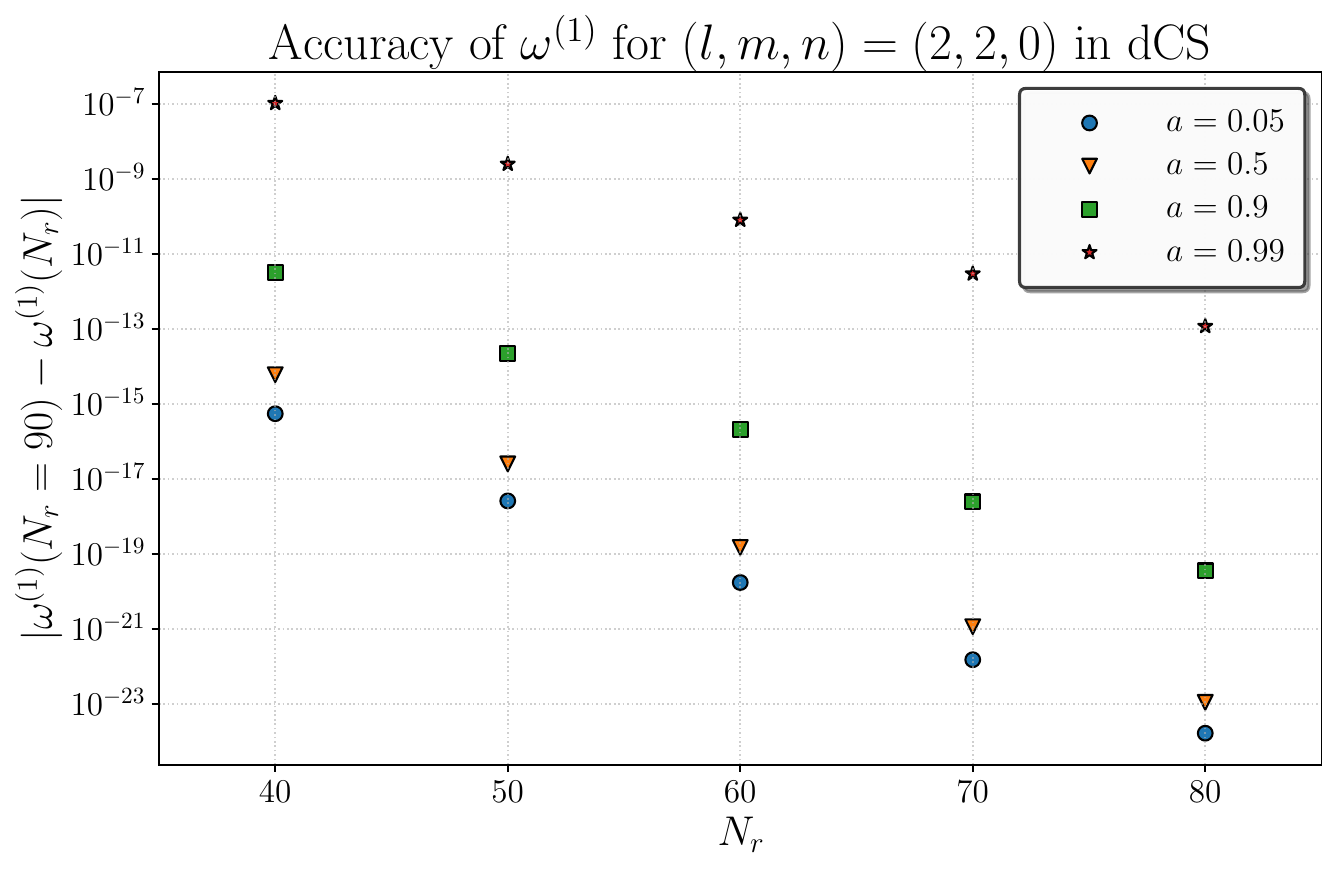}
    \label{fig:l2convergencedCS}
    }%
    \caption{Convergence of lowest order correction to $l=m=2$ scalar QNMs as a function of radial grid size. Blue circles represent convergence of the $a=0.05$ case, orange triangles $a=0.5,$ green squares $a=0.9$ and purple stars $a=0.99.$ The corresponding angular grid size of the computations is given by $N_y=N_r-20$.}
    \label{fig:convergence}
\end{figure*}

\subsection{Dynamical Chern-Simons}

The leading order corrections $\omega^{(1)}$ for the $l=m=0$ mode in dCS gravity are shown in Fig.~\ref{fig:l0dCS}. As in the sGB case, we compare our results obtained using the spectral background solutions with those derived from the spin expansion background of \cite{canoLeadingHigherderivativeCorrections2019}. At moderate spins the two approaches yield consistent results, while noticeable deviations appear for larger spins where the spin expansion becomes less accurate. The corrections to the $l=m=0$ mode frequency bend around in the complex plane, which is a trend we do not observe using the analytical background based on a spin expansion. Excellent agreement with the EVP method is once again recovered in the low-spin regime.

A key difference with respect to sGB arises in the Schwarzschild limit. In dCS gravity the scalar sector does not receive corrections in the non-rotating case due to the parity structure of the theory. Consequently, the scalar QNM corrections vanish at $a=0$, which is reflected in Figs.~\ref{fig:l0dCS} and \ref{fig:l2dCS} by the fact that the trajectories of $\omega^{(1)}$ originate at the origin in the complex plane.

The behaviour of the higher multipoles follows a pattern similar to the sGB case. As an example, the $l=2$ corrections are displayed in Fig.~\ref{fig:l2dCS}. For several modes, most prominently the $l=m=2$ mode, the corrections grow rapidly as the spin increases when $a>0.9$. Furthermore, we have large corrections for $l=1$ at $m=1$, for $l=3$ at $m=2,3$, for $l=4$ at $m=3,4$, and for $l=5$ at $m=3,4,5$. This is the same list of modes as observed in sGB, suggesting this behaviour is theory independent. 

To verify the robustness of our results, we have also examined the numerical convergence of the solutions. As shown in Fig.~\ref{fig:l2convergencedCS}, the corrections for the fundamental $l=m=2$ mode converge exponentially with increasing grid resolution, as expected for pseudo-spectral methods. As in the sGB case, higher spins require slightly larger grids to reach the same accuracy, whereas higher multipoles converge more rapidly.

Overall, the frequency shifts are obtained with relative errors smaller than $10^{-3}$ for all spins and modes considered. For $l>1$ the relative errors are typically below $10^{-6}$. We have additionally verified that the resolution of the background spacetime has a negligible impact on the results: reducing the spectral order of the background produces changes in the QNM corrections that are well below the numerical error of the pseudo-spectral solver. This indicates that, as in the sGB case, the dominant source of numerical uncertainty arises from the resolution of the QNM grid rather than the background geometry.

\subsection{Large Spin Behaviour}\label{sec:divergent}

\begin{figure}[b]
\centering
\includegraphics[width=\linewidth]{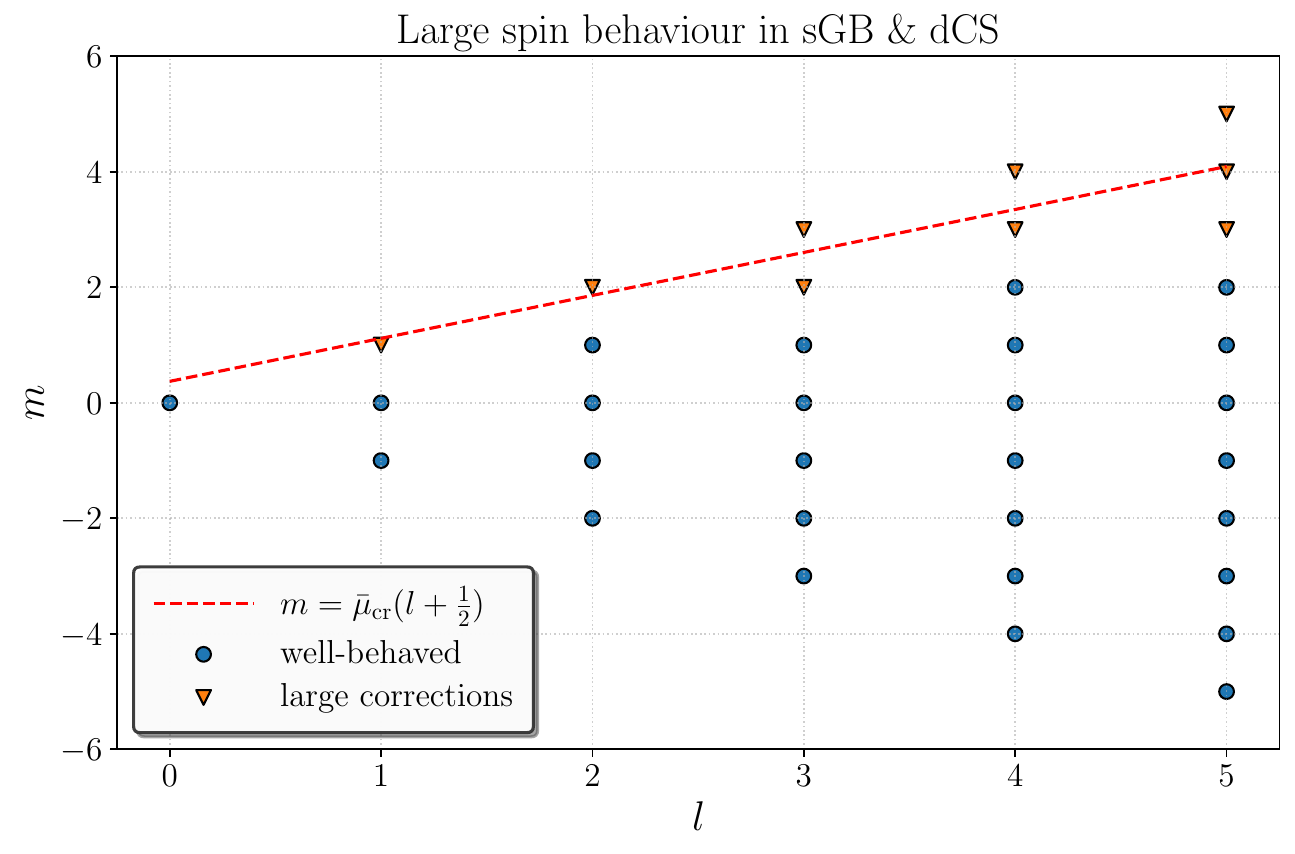}
\caption{Phase space diagram to illustrate which modes seem to diverge in the large-spin limit and which remain well-behaved, along with the DM-ZDM phase boundary in the Eikonal limit (red dashed line). Notably, the modes that exhibit rapid growth in the large-spin regime are the same for sGB and dCS}
\label{fig:phase_sGB}
\vspace{-2\baselineskip}
\end{figure}

We have observed that certain modes grow by orders of magnitude for large spins and exhibit apparent divergence. This subset of divergent modes is the same for sGB and dCS. We have summarized these results in Fig.~\ref{fig:phase_sGB}, where we show which modes are well-behaved and which modes receive large corrections. Similar to findings in \cite{canoAmplificationNewPhysics2025}, we observe that the modes receiving large corrections are close to the phase boundary, which separates the regions for near-extremal BHs where only damped modes (DMs) exist from the one where zero damped modes (ZDMs) and DMs coexist. For a Kerr BH in GR, this phase boundary is, in the Eikonal limit $l\gg1$, given by $m=\bar{\mu}_\mathrm{cr}(l+\frac{1}{2})$ with $\bar{\mu}_\mathrm{cr}\approx0.744$ \cite{yangBranchingQuasinormalModes2013, yangQuasinormalModesNearly2013}. 

Assuming that the lowest-order approximation remains valid and that our limited spin range $a\leq0.99$ sufficiently probes the near-extremal regime, modes near the phase boundary acquire stronger damping ($\Im(\omega^{(1)})<0$) for a given $l$ , effectively pushing the zero-damped boundary to larger $m$. This, in turn, implies that $\bar{\mu}_\mathrm{cr}$ should receive positive corrections in the theories we considered. However, as we will discuss in Sec.~\ref{sec:disc}, there are concerns about the validity of these two assumptions.

Furthermore, in both sGB and dCS we have found corrections $\omega^{(1)}$ for which $\Im[\omega^{(1)}]>0$. These corrections may indicate potential linear instabilities, since there exist values of $\lambda$ for which $\Im[\omega^{(0)}+\lambda\omega^{(1)}]>0$ where modes become undamped. Using the largest correction at $a=0.98$ for $l=m=0$, we find that undamped modes occur in lowest-order for $\lambda\gtrsim 0.81$ in sGB and $\lambda\gtrsim 0.52$ in dCS. These values are well above current observational constraints. In particular, for shift-symmetric sGB, the bound $\lambda \le 0.019$ is obtained by comparing GW230529 to post-Newtonian predictions, using the BH mass of $3.6\:\text{M}_\odot$ for this BH-NS merger \cite{sangerTestsGeneralRelativity2024}. For dCS, a bound of $\lambda\le 0.23$ is obtained from independent measurements of neutron star properties \cite{silvaAstrophysicalTheoreticalPhysics2021}. Therefore, our results do not indicate the presence of linear instabilities within the observationally allowed parameter space.

\section{Conclusion and Discussion}\label{sec:disc}

We have obtained the spectrum for scalar QNMs in sGB and dCS up to a spin of $a/M=0.99$, using a pseudo-spectral QNM solver applied to spectral solutions for the background. The approach presented here is general and can be applied to other higher-curvature theories (see \cite{vandersteenRingingRapidlyRotating2026} for the cubic higher-derivative corrections). For all spins considered, we obtain reliable frequency shifts with exponential convergence, as expected for pseudo-spectral schemes. As can be seen in Fig.~\ref{fig:l0sGB} and Fig.~\ref{fig:l0dCS}, spin expansions of the background metric cease to be reliable at high spins, highlighting the importance of using fully numerical background solutions when approaching the large spin regime.

Our results show that the beyond-GR corrections $\omega^{(1)}$ generally increase with spin throughout the astrophysically relevant range. This behaviour supports the idea that higher-derivative gravity effects are amplified for rapidly rotating BHs, making them prime objects to test for deviations beyond GR \cite{paniScalarShortcutBeyondKerr2026}.

Although the behaviour observed here indicates a breakdown of the leading-order expansion in the coupling parameter, this does not necessarily imply a breakdown of the EFT itself. The EFT fails only when higher-order operators in the effective Lagrangian become non-negligible. It is therefore possible that the theory remains within its regime of validity even when the lowest-order perturbative approximation to the QNM spectrum in Eqs.~\eqref{eq:psi_expanded}-\eqref{eq:om_expanded} becomes unreliable. Therefore these results provide no concrete evidence of a breakdown of the EFT.

As discussed in Sec.~\ref{sec:divergent}, the large corrections observed for certain modes at high spin can be understood in terms of a shift in the phase boundary between two classes of Kerr QNMs: ZDMs, which become infinitely long-lived in the extremal limit, and DMs, which retain finite damping \cite{canoAmplificationNewPhysics2025, yangBranchingQuasinormalModes2013, yangQuasinormalModesNearly2013}. In beyond-GR theories, this boundary also receives corrections and its location shifts in the parameter space. As a consequence, modes that lie in the ZDM+DM regime in GR may move into the DM regime in modified gravity, or vice versa. The former is suggested to happen based on the spin range and lowest-order corrections probed in this study. This order-one change in the nature of the mode leads to corrections that appear to diverge when computed as a first-order perturbation around the GR spectrum, since one has to simultaneously expand in the beyond GR-coupling as well as in the distance from the phase boundary. The first-order expansion around the GR spectrum breaks down when the beyond-GR shift in the phase boundary becomes comparable to the distance from that boundary. This effect is most pronounced for modes with $l\approx m$, which lie closest to the ZDM–DM boundary, while modes with negative $m$, located far from this boundary, show no such behaviour. This is in agreement with our findings.

For spins that remain a finite distance from extremality the corrections remain well-behaved. However, as the extremal limit is approached, the domain of validity of the leading-order expansion shrinks for modes sufficiently close to the phase boundary. This accounts for the rapid growth of the corrections observed in our numerical results and suggests that the perturbative expansion in the coupling parameter $\lambda$ breaks down in the near-extremal regime, where an additional expansion in the distance to the phase boundary becomes necessary.

We also note that the extremality condition itself receives corrections at the same order in $\lambda$. Extremality is defined by the vanishing of the surface gravity $\kappa=0$, which no longer corresponds to $a=1$ in higher-curvature theories. Previous work in \cite{chungQuasinormalModeFrequencies2024} has shown that the leading correction to the surface gravity shifts the extremality condition to $a>1$ for $\lambda=\alpha^2/M^4>0$. As a consequence, the true extremal solution lies beyond the parameter range currently accessible with numerical methods and backgrounds.

Our numerical results complement earlier theoretical work and support the observation that higher-derivative effects are amplified in the near-extremal limit \cite{canoEikonalQuasinormalModes2025, canoAmplificationNewPhysics2025}. Roughly the same orders of magnitude of amplification have been found, despite working with different theories. This strengthens the claim that the phase boundary shifts act as a theory-independent amplification of beyond GR effects in the near-extremal limit. This suggests that rapidly rotating BHs may offer the most promising observational prospects for detecting or constraining higher-derivative corrections with gravitational-wave measurements, because the phase boundary effect is spin-weight independent. Future work could extend these calculations to spins even closer to extremality. As suggested by Fig.~1 of \cite{canoEikonalQuasinormalModes2025}, it is possible that modes which appear to diverge at moderately high spins may eventually level off once the system is examined sufficiently close to the extremal limit, provided the modes lie far enough from the shifted phase boundary.

\section{Acknowledgements}

We thank Llibert Aresté Saló, Pablo Cano, Vitor Cardoso, Marina David, Pedro Fernandes, Nicola Franchini, and Henri Inchausp\'e for useful discussions.
T.v.d.S. acknowledges funding from the Research Foundation - Flanders (FWO, Fonds Wetenschappelijk Onderzoek) through a PhD Fellowship under FWO Grant No. 1125026N.
T.v.d.S. and T.H. thank the Belgian Federal Science Policy Office (BELSPO) for the provision of financial support in the framework of the PRODEX Programme of the European Space Agency (ESA) under contract number PEA4000144253.
This research was supported in part by KU Leuven grant C16/25/010.
S.M.~acknowledges funding from the Simons Foundation International and the Simons Foundation through Grant No. SFI-MPS-BH-00012593-11.
We acknowledge support by the VILLUM Foundation (Grant No. VIL37766).
The Center of Gravity is a Center of Excellence funded by the Danish National Research Foundation under Grant No. DNRF184.
K.K.H.L~and N.Y.~acknowledge support from the Simons Foundation through Award No. 896696, the Simons Foundation International through Award No. SFI-MPS-BH-00012593-01, the NSF through Grants No. PHY-2207650 and PHY-25-12423, and NASA through Grant No. 80NSSC22K0806.
A.K.W.C~acknowledges support from the Herchel Smith Fellowship at the University of Cambridge.
T.G.F.L.~is supported by grants from the the Research Foundation – Flanders (FWO) Grant No. I002123N and I000725N.

\bibliography{bibliography}

@article{alexanderChernSimonsModified2009,
  title = {Chern-{{Simons Modified General Relativity}}},
  author = {Alexander, Stephon and Yunes, Nicol{\'a}s},
  year = 2009,
  journal = {Physics Reports},
  volume = {480},
  number = {1},
  eprint = {0907.2562},
  primaryclass = {hep-th},
  pages = {1--55},
  issn = {0370-1573},
  doi = {10.1016/j.physrep.2009.07.002},
  urldate = {2026-03-10},
  archiveprefix = {arXiv}
}

@misc{alexanderCosmologyGraviAxions2025,
  title = {Cosmology of {{Gravi-Axions}}},
  author = {Alexander, Stephon and Gabadadze, Gregory and Jenks, Leah and Yunes, Nicol{\'a}s},
  year = 2025,
  month = nov,
  eprint = {2512.00154},
  primaryclass = {hep-ph},
  publisher = {arXiv},
  doi = {10.48550/arXiv.2512.00154},
  urldate = {2026-03-19},
  archiveprefix = {arXiv}
}

@article{ayzenbergSlowlyRotatingBlack2014,
  title = {Slowly Rotating Black Holes in {{Einstein-Dilaton-Gauss-Bonnet}} Gravity: {{Quadratic}} Order in Spin Solutions},
  shorttitle = {Slowly Rotating Black Holes in {{Einstein-Dilaton-Gauss-Bonnet}} Gravity},
  author = {Ayzenberg, Dimitry and Yunes, Nicol{\'a}s},
  year = 2014,
  month = aug,
  journal = {Physical Review D},
  volume = {90},
  number = {4},
  eprint = {1405.2133},
  primaryclass = {gr-qc},
  pages = {044066},
  publisher = {American Physical Society},
  doi = {10.1103/PhysRevD.90.044066},
  urldate = {2026-03-11},
  archiveprefix = {arXiv}
}

@article{bergshoeffTendimensionalMaxwellEinsteinSupergravity1982,
  title = {Ten-{{Dimensional Maxwell-Einstein Supergravity}}, {{Its Currents}}, and the {{Issue}} of {{Its Auxiliary Fields}}},
  author = {Bergshoeff, E. and {de Roo}, M. and {de Wit}, B. and {van Nieuwenhuizen}, P.},
  year = 1982,
  journal = {Nuclear Physics B},
  volume = {195},
  number = {1},
  pages = {97--136},
  issn = {0550-3213},
  doi = {10.1016/0550-3213(82)90050-5},
  urldate = {2026-03-10}
}

@article{bertiBlackHoleSpectroscopy2025,
  author={Berti, Emanuele and Cardoso, Vitor and Carullo, Gregorio and others},
  title={Black hole spectroscopy: from theory to experiment},
  journal={Classical and Quantum Gravity},
  url={http://iopscience.iop.org/article/10.1088/1361-6382/ae59e2},
  year={2026},
  shorttitle = {Black Hole Spectroscopy},
  editor = {Berti, Emanuele and Cardoso, Vitor and Carullo, Gregorio},
  eprint = {2505.23895},
  primaryclass = {gr-qc},
  publisher = {IOP Publishing},
  archiveprefix = {arXiv}
}

@article{bertiExtremeGravityTests2018,
  title = {Extreme {{Gravity Tests}} with {{Gravitational Waves}} from {{Compact Binary Coalescences}}: ({{II}}) {{Ringdown}}},
  shorttitle = {Extreme Gravity Tests with Gravitational Waves from Compact Binary Coalescences},
  author = {Berti, Emanuele and Yagi, Kent and Yang, Huan and Yunes, Nicol{\'a}s},
  year = 2018,
  month = apr,
  journal = {General Relativity and Gravitation},
  volume = {50},
  number = {5},
  eprint = {1801.03587},
  primaryclass = {gr-qc},
  pages = {49},
  issn = {1572-9532},
  doi = {10.1007/s10714-018-2372-6},
  urldate = {2026-02-26},
  archiveprefix = {arXiv},
  langid = {english}
}

@article{bertiGravitationalwaveSpectroscopyMassive2006,
  title = {On Gravitational-Wave Spectroscopy of Massive Black Holes with the Space Interferometer {{LISA}}},
  author = {Berti, Emanuele and Cardoso, Vitor and Will, Clifford M.},
  year = 2006,
  journal = {Physical Review D},
  volume = {73},
  number = {6},
  eprint = {gr-qc/0512160},
  pages = {064030},
  publisher = {American Physical Society},
  doi = {10.1103/PhysRevD.73.064030},
  urldate = {2025-08-01},
  archiveprefix = {arXiv}
}

@article{bertiQuasinormalModesBlackholes2009,
  title = {Quasinormal Modes of Black Holes and Black Branes},
  author = {Berti, Emanuele and Cardoso, Vitor and Starinets, Andrei O.},
  year = 2009,
  journal = {Class. Quant. Grav.},
  volume = {26},
  eprint = {0905.2975},
  primaryclass = {gr-qc},
  pages = {163001},
  doi = {10.1088/0264-9381/26/16/163001},
  archiveprefix = {arXiv}
}

@article{blazquez-salcedoQuasinormalModesKerr2024,
  title = {Quasinormal Modes of {{Kerr}} Black Holes Using a Spectral Decomposition of the Metric Perturbations},
  author = {{Bl{\'a}zquez-Salcedo}, Jose Luis and Khoo, Fech Scen and Kunz, Jutta and {Gonz{\'a}lez-Romero}, Luis Manuel},
  year = 2024,
  month = mar,
  journal = {Physical Review D},
  volume = {109},
  number = {6},
  eprint = {2312.10754},
  primaryclass = {gr-qc},
  pages = {064028},
  publisher = {American Physical Society},
  doi = {10.1103/PhysRevD.109.064028},
  urldate = {2026-03-10},
  archiveprefix = {arXiv}
}

@article{blazquezsalcedoPerturbedBlackHoles2016,
  title = {Perturbed Black Holes in {{Einstein-dilaton-Gauss-Bonnet}} Gravity: {{Stability}}, Ringdown, and Gravitational-Wave Emission},
  author = {{Bl{\'a}zquez-Salcedo}, Jose Luis and Macedo, Caio F. B. and Cardoso, Vitor and Ferrari, Valeria and Gualtieri, Leonardo and Khoo, Fech Scen and Kunz, Jutta and Pani, Paolo},
  year = 2016,
  month = nov,
  journal = {Physical Review D: Particles and Fields},
  volume = {94},
  number = {10},
  eprint = {1609.01286},
  primaryclass = {gr-qc},
  pages = {104024},
  doi = {10.1103/PhysRevD.94.104024},
  archiveprefix = {arXiv}
}

@article{blazquezsalcedoQuasinormalmodesEinsteingaussbonnetdilatonBlackholes2017,
  title = {Quasinormal Modes of {{Einstein-Gauss-Bonnet-dilaton}} Black Holes},
  author = {{Bl{\'a}zquez-Salcedo}, Jose Luis and Khoo, Fech Scen and Kunz, Jutta},
  year = 2017,
  month = sep,
  journal = {Physical Review D: Particles and Fields},
  volume = {96},
  number = {6},
  eprint = {1706.03262},
  primaryclass = {gr-qc},
  pages = {064008},
  publisher = {American Physical Society},
  doi = {10.1103/PhysRevD.96.064008},
  archiveprefix = {arXiv}
}

@misc{boyceKerrBlackHole2026,
  title = {Kerr {{Black Hole Ringdown}} in {{Effective Field Theory}}},
  author = {Boyce, William L. and Santos, Jorge E.},
  year = 2026,
  month = mar,
  eprint = {2603.10102},
  primaryclass = {gr-qc},
  publisher = {arXiv},
  doi = {10.48550/arXiv.2603.10102},
  urldate = {2026-03-12},
  archiveprefix = {arXiv}
}

@book{boydChebyshevFourierSpectral2001,
  title = {Chebyshev and {{Fourier}} Spectral Methods},
  author = {Boyd, John P.},
  year = 2001,
  series = {Dover Books on Mathematics},
  edition = {1. publ., rev. and enlarged 2. ed. of the work by Springer, Berlin, 1969},
  publisher = {Dover Publ},
  address = {Mineola, NY},
  isbn = {978-0-486-41183-5},
  langid = {english}
}

@article{CanoAccuracySlowrotationApproximation2024,
  title = {Accuracy of the Slow-Rotation Approximation for Black Holes in Modified Gravity in Light of Astrophysical Observables},
  author = {Cano, Pablo A. and Deich, Alexander and Yunes, Nicol{\'a}s},
  year = 2024,
  month = jan,
  journal = {Physical Review D: Particles and Fields},
  volume = {109},
  number = {2},
  eprint = {2305.15341},
  primaryclass = {gr-qc},
  pages = {024048},
  doi = {10.1103/PhysRevD.109.024048},
  archiveprefix = {arXiv}
}

@misc{canoAmplificationNewPhysics2025,
  title = {Amplification of New Physics in the Quasinormal Mode Spectrum of Highly-Rotating Black Holes},
  author = {Cano, Pablo A. and David, Marina and {van der Velde}, Guido},
  year = 2025,
  month = oct,
  eprint = {2510.17962},
  primaryclass = {gr-qc},
  publisher = {arXiv},
  doi = {10.48550/arXiv.2510.17962},
  urldate = {2025-11-05},
  archiveprefix = {arXiv}
}

@misc{canoEikonalQuasinormalModes2025,
  title = {Eikonal Quasinormal Modes of Highly-Spinning Black Holes in Higher-Curvature Gravity: A Window into Extremality},
  shorttitle = {Eikonal Quasinormal Modes of Highly-Spinning Black Holes in Higher-Curvature Gravity},
  author = {Cano, Pablo A. and David, Marina and {van der Velde}, Guido},
  year = 2025,
  month = sep,
  eprint = {2509.08664},
  primaryclass = {gr-qc},
  publisher = {arXiv},
  doi = {10.48550/arXiv.2509.08664},
  urldate = {2025-09-15},
  archiveprefix = {arXiv}
}

@article{canoGravitationalRingingRotating2022,
  title = {Gravitational Ringing of Rotating Black Holes in Higher-Derivative Gravity},
  author = {Cano, Pablo A. and Fransen, Kwinten and Hertog, Thomas and Maenaut, Simon},
  year = 2022,
  month = jan,
  journal = {Physical Review D: Particles and Fields},
  volume = {105},
  number = {2},
  eprint = {2110.11378},
  primaryclass = {gr-qc},
  pages = {024064},
  doi = {10.1103/PhysRevD.105.024064},
  archiveprefix = {arXiv}
}

@article{canoHigherderivativeCorrectionsKerr2024,
  title = {Higher-Derivative Corrections to the {{Kerr}} Quasinormal Mode Spectrum},
  author = {Cano, Pablo A. and Capuano, Lodovico and Franchini, Nicola and Maenaut, Simon and V{\"o}lkel, Sebastian H.},
  year = 2024,
  month = dec,
  journal = {Physical Review D: Particles and Fields},
  volume = {110},
  number = {12},
  eprint = {2409.04517},
  primaryclass = {gr-qc},
  pages = {124057},
  doi = {10.1103/PhysRevD.110.124057},
  archiveprefix = {arXiv}
}

@article{canoLeadingHigherderivativeCorrections2019,
  title = {Leading Higher-Derivative Corrections to {{Kerr}} Geometry},
  author = {Cano, Pablo A. and Ruip{\'e}rez, Alejandro},
  year = 2019,
  month = may,
  journal = {Journal of High Energy Physics},
  volume = {2019},
  number = {5},
  eprint = {1901.01315},
  primaryclass = {gr-qc},
  pages = {189},
  issn = {1029-8479},
  doi = {10.1007/JHEP05(2019)189},
  urldate = {2025-01-13},
  archiveprefix = {arXiv},
  langid = {english}
}

@article{canoParametrizedQuasinormalMode2024,
  title = {Parametrized Quasinormal Mode Framework for Modified {{Teukolsky}} Equations},
  author = {Cano, Pablo A. and Capuano, Lodovico and Franchini, Nicola and Maenaut, Simon and V{\"o}lkel, Sebastian H.},
  year = 2024,
  month = nov,
  journal = {Physical Review D},
  volume = {110},
  number = {10},
  eprint = {2407.15947},
  primaryclass = {gr-qc},
  pages = {104007},
  publisher = {American Physical Society},
  doi = {10.1103/PhysRevD.110.104007},
  urldate = {2025-01-15},
  archiveprefix = {arXiv}
}

@article{canoQuasinormalModesRotating2023,
  title = {Quasinormal Modes of Rotating Black Holes in Higher-Derivative Gravity},
  author = {Cano, Pablo A. and Fransen, Kwinten and Hertog, Thomas and Maenaut, Simon},
  year = 2023,
  month = dec,
  journal = {Physical Review D: Particles and Fields},
  volume = {108},
  number = {12},
  eprint = {2307.07431},
  primaryclass = {gr-qc},
  pages = {124032},
  doi = {10.1103/PhysRevD.108.124032},
  archiveprefix = {arXiv}
}

@article{canoRingingRotatingBlack2020,
  title = {Ringing of Rotating Black Holes in Higher-Derivative Gravity},
  author = {Cano, Pablo A. and Fransen, Kwinten and Hertog, Thomas},
  year = 2020,
  month = aug,
  journal = {Physical Review D},
  volume = {102},
  number = {4},
  eprint = {2005.03671},
  primaryclass = {gr-qc},
  pages = {044047},
  issn = {2470-0010, 2470-0029},
  doi = {10.1103/PhysRevD.102.044047},
  urldate = {2025-01-15},
  archiveprefix = {arXiv},
  langid = {english}
}

@article{canoTeukolskyEquationNearextremal2024,
  title = {Teukolsky Equation for Near-Extremal Black Holes beyond General Relativity: {{Near-horizon}} Analysis},
  author = {Cano, Pablo A. and David, Marina},
  year = 2024,
  month = sep,
  journal = {Physical Review D: Particles and Fields},
  volume = {110},
  number = {6},
  eprint = {2407.02017},
  primaryclass = {gr-qc},
  pages = {064067},
  doi = {10.1103/PhysRevD.110.064067},
  archiveprefix = {arXiv}
}

@article{canoUniversalTeukolskyEquations2023,
  title = {Universal {{Teukolsky}} Equations and Black Hole Perturbations in Higher-Derivative Gravity},
  author = {Cano, Pablo A. and Fransen, Kwinten and Hertog, Thomas and Maenaut, Simon},
  year = 2023,
  month = jul,
  journal = {Physical Review D},
  volume = {108},
  number = {2},
  eprint = {2304.02663},
  primaryclass = {gr-qc},
  pages = {024040},
  publisher = {American Physical Society},
  doi = {10.1103/PhysRevD.108.024040},
  urldate = {2024-09-20},
  archiveprefix = {arXiv}
}

@book{canutoSpectralMethodsFundamentals2006,
  title = {Spectral {{Methods}}: {{Fundamentals}} in {{Single Domains}}},
  shorttitle = {Spectral {{Methods}}},
  author = {Canuto, Claudio and Hussaini, M. Youssuff and Quarteroni, Alfio and Zang, Thomas A.},
  year = 2006,
  series = {Scientific {{Computation}}},
  publisher = {Springer},
  address = {Berlin, Heidelberg},
  doi = {10.1007/978-3-540-30726-6},
  urldate = {2026-02-13},
  isbn = {978-3-540-30725-9 978-3-540-30726-6},
  langid = {english}
}

@misc{vandersteenRingingRapidlyRotating2026,
  title = {Ringing of Rapidly Rotating Black Holes in Effective Field Theory},
  author = {{van der Steen}, Tom and Maenaut, Simon and Husken, Stef J.B. and Fernandes, Pedro G.S. and Jockwer, Maxim D. and Cardoso, Vitor and Hertog, Thomas and Li, Tjonnie G.F.},
  year = 2026,
  month = apr,
  eprint = {2604.11755},
  primaryclass = {gr-qc},
  publisher = {arXiv},
  doi = {10.48550/arXiv.2604.11755},
  url = {http://arxiv.org/abs/2604.11755},
  urldate = {2026-04-14},
  archiveprefix = {arXiv},
}

@article{cardosoBlackHolesEffective2018,
  title = {Black {{Holes}} in an {{Effective Field Theory Extension}} of {{General Relativity}}},
  author = {Cardoso, Vitor and Kimura, Masashi and Maselli, Andrea and Senatore, Leonardo},
  year = 2018,
  month = dec,
  journal = {Physical Review Letters},
  volume = {121},
  number = {25},
  eprint = {1808.08962},
  primaryclass = {gr-qc},
  pages = {251105},
  publisher = {American Physical Society},
  doi = {10.1103/PhysRevLett.121.251105},
  urldate = {2026-03-11},
  archiveprefix = {arXiv}
}

@article{cardosoParametrizedBlackholeQuasinormal2019,
  title = {Parametrized Black Hole Quasinormal Ringdown: {{Decoupled}} Equations for Nonrotating Black Holes},
  author = {Cardoso, Vitor and Kimura, Masashi and Maselli, Andrea and Berti, Emanuele and Macedo, Caio F.B. and McManus, Ryan},
  year = 2019,
  month = may,
  journal = {Physical Review D: Particles and Fields},
  volume = {99},
  number = {10},
  eprint = {1901.01265},
  primaryclass = {gr-qc},
  pages = {104077},
  publisher = {American Physical Society},
  doi = {10.1103/PhysRevD.99.104077},
  archiveprefix = {arXiv}
}

@article{cardosoPerturbationsSchwarzschildBlakcholes2009,
  title = {Perturbations of {{Schwarzschild}} Black Holes in {{Dynamical Chern-Simons}} Modified Gravity},
  author = {Cardoso, Vitor and Gualtieri, Leonardo},
  year = 2009,
  journal = {Physical Review D: Particles and Fields},
  volume = {80},
  eprint = {0907.5008},
  primaryclass = {gr-qc},
  pages = {064008},
  doi = {10.1103/PhysRevD.81.089903},
  archiveprefix = {arXiv}
}

@article{carulloEmpiricalTestsBlack2018,
  title = {Empirical Tests of the Black Hole No-Hair Conjecture Using Gravitational-Wave Observations},
  author = {Carullo, Gregorio and Van Der Schaaf, Laura and London, Lionel and others},
  year = 2018,
  month = nov,
  journal = {Physical Review D},
  volume = {98},
  number = {10},
  eprint = {1805.04760},
  primaryclass = {gr-qc},
  pages = {104020},
  publisher = {American Physical Society},
  doi = {10.1103/PhysRevD.98.104020},
  urldate = {2026-02-26},
  archiveprefix = {arXiv}
}

@article{carulloEnhancingModifiedGravity2021,
  title = {Enhancing Modified Gravity Detection from Gravitational-Wave Observations Using the Parametrized Ringdown Spin Expansion Coeffcients Formalism},
  author = {Carullo, Gregorio},
  year = 2021,
  month = jun,
  journal = {Physical Review D: Particles and Fields},
  volume = {103},
  number = {12},
  eprint = {2102.05939},
  primaryclass = {gr-qc},
  pages = {124043},
  doi = {10.1103/PhysRevD.103.124043},
  archiveprefix = {arXiv}
}

@misc{chungProbingQuadraticGravity2025a,
  title = {Probing Quadratic Gravity with Black-Hole Ringdown Gravitational Waves Measured by {{LIGO-Virgo-KAGRA}} Detectors},
  author = {Chung, Adrian Ka-Wai and Yunes, Nicol{\'a}s},
  year = 2025,
  month = jun,
  eprint = {2506.14695},
  primaryclass = {gr-qc},
  publisher = {arXiv},
  doi = {10.48550/arXiv.2506.14695},
  urldate = {2026-03-12},
  archiveprefix = {arXiv}
}

@article{chungQuasinormalModeFrequencies2024,
  title = {Quasinormal Mode Frequencies and Gravitational Perturbations of Black Holes with Any Subextremal Spin in Modified Gravity through {{METRICS}}: {{The}} Scalar-{{Gauss-Bonnet}} Gravity Case},
  shorttitle = {Quasinormal Mode Frequencies and Gravitational Perturbations of Black Holes with Any Subextremal Spin in Modified Gravity through {{METRICS}}},
  author = {Chung, Adrian Ka-Wai and Yunes, Nicolas},
  year = 2024,
  month = sep,
  journal = {Physical Review D},
  volume = {110},
  number = {6},
  eprint = {2406.11986},
  primaryclass = {gr-qc},
  pages = {064019},
  doi = {10.1103/PhysRevD.110.064019},
  archiveprefix = {arXiv}
}

@article{chungQuasinormalModeFrequencies2025,
  title = {Quasinormal Mode Frequencies and Gravitational Perturbations of Spinning Black Holes in Modified Gravity through {{METRICS}}: {{The}} Dynamical {{Chern-Simons}} Gravity Case},
  shorttitle = {Quasinormal Mode Frequencies and Gravitational Perturbations of Spinning Black Holes in Modified Gravity through {{METRICS}}},
  author = {Chung, Adrian Ka-Wai and Lam, Kelvin Ka-Ho and Yunes, Nicolas},
  year = 2025,
  month = jun,
  journal = {Physical Review D},
  volume = {111},
  number = {12},
  eprint = {2503.11759},
  primaryclass = {gr-qc},
  pages = {124052},
  issn = {2470-0010, 2470-0029},
  doi = {10.1103/g83n-rrlj},
  urldate = {2026-03-02},
  archiveprefix = {arXiv}
}

@article{chungRingingOutGeneral2024,
  title = {Ringing {{Out General Relativity}}: {{Quasinormal Mode Frequencies}} for {{Black Holes}} of {{Any Spin}} in {{Modified Gravity}}},
  shorttitle = {Ringing out {{General Relativity}}},
  author = {Chung, Adrian Ka-Wai and Yunes, Nicolas},
  year = 2024,
  month = oct,
  journal = {Physical Review Letters},
  volume = {133},
  number = {18},
  eprint = {2405.12280},
  primaryclass = {gr-qc},
  pages = {181401},
  issn = {0031-9007, 1079-7114},
  doi = {10.1103/PhysRevLett.133.181401},
  urldate = {2026-03-20},
  archiveprefix = {arXiv}
}

@article{chungSpectralMethodGravitational2023,
  title = {Spectral Method for the Gravitational Perturbations of Black Holes: {{Schwarzschild}} Background Case},
  shorttitle = {Spectral {{Method}} for the {{Gravitational Perturbations}} of {{Black Holes}}},
  author = {Chung, Adrian Ka-Wai and Wagle, Pratik and Yunes, Nicolas},
  year = 2023,
  month = jun,
  journal = {Physical Review D},
  volume = {107},
  number = {12},
  eprint = {2302.11624},
  primaryclass = {gr-qc},
  pages = {124032},
  issn = {2470-0010, 2470-0029},
  doi = {10.1103/PhysRevD.107.124032},
  urldate = {2026-03-20},
  archiveprefix = {arXiv}
}

@article{chungSpectralMethodMetric2024,
  title = {Spectral Method for Metric Perturbations of Black Holes: {{Kerr}} Background Case in General Relativity},
  shorttitle = {Spectral Method for Metric Perturbations of Black Holes},
  author = {Chung, Adrian Ka-Wai and Wagle, Pratik and Yunes, Nicolas},
  year = 2024,
  month = feb,
  journal = {Physical Review D},
  volume = {109},
  number = {4},
  eprint = {2312.08435},
  primaryclass = {gr-qc},
  pages = {044072},
  issn = {2470-0010, 2470-0029},
  doi = {10.1103/PhysRevD.109.044072},
  urldate = {2026-03-20},
  archiveprefix = {arXiv}
}

@article{cookGravitationalPerturbationsKerr2014,
  title = {Gravitational Perturbations of the {{Kerr}} Geometry: {{High-accuracy}} Study},
  shorttitle = {Gravitational Perturbations of the {{Kerr}} Geometry},
  author = {Cook, Gregory B. and Zalutskiy, Maxim},
  year = 2014,
  month = dec,
  journal = {Physical Review D},
  volume = {90},
  number = {12},
  eprint = {1410.7698},
  primaryclass = {gr-qc},
  pages = {124021},
  publisher = {American Physical Society},
  doi = {10.1103/PhysRevD.90.124021},
  urldate = {2025-01-13},
  archiveprefix = {arXiv}
}

@article{derhamBlackholeGravitationalWaves2020,
  title = {Black {{Hole Gravitational Waves}} in the {{Effective Field Theory}} of {{Gravity}}},
  author = {{de Rham}, Claudia and Francfort, J{\'e}r{\'e}mie and Zhang, Jun},
  year = 2020,
  month = jul,
  journal = {Physical Review D: Particles and Fields},
  volume = {102},
  number = {2},
  eprint = {2005.13923},
  primaryclass = {hep-th},
  pages = {024079},
  publisher = {American Physical Society},
  doi = {10.1103/PhysRevD.102.024079},
  archiveprefix = {arXiv}
}

@article{diasNumericalMethodsFinding2016,
  title = {Numerical {{Methods}} for {{Finding Stationary Gravitational Solutions}}},
  author = {Dias, {\'O}scar J.C. and Santos, Jorge E. and Way, Benson},
  year = 2016,
  month = jun,
  journal = {Classical and Quantum Gravity},
  volume = {33},
  number = {13},
  eprint = {1510.02804},
  primaryclass = {hep-th},
  pages = {133001},
  publisher = {IOP Publishing},
  issn = {0264-9381},
  doi = {10.1088/0264-9381/33/13/133001},
  urldate = {2025-02-04},
  archiveprefix = {arXiv},
  langid = {english}
}

@article{donoghueGeneralRelativityEffective1994,
  title = {General Relativity as an Effective Field Theory: {{The}} Leading Quantum Corrections},
  shorttitle = {General Relativity as an Effective Field Theory},
  author = {Donoghue, John F.},
  year = 1994,
  journal = {Physical Review D},
  volume = {50},
  number = {6},
  eprint = {gr-qc/9405057},
  pages = {3874--3888},
  issn = {0556-2821},
  doi = {10.1103/PhysRevD.50.3874},
  urldate = {2026-03-10},
  archiveprefix = {arXiv}
}

@article{dreyerBlackholeSpectroscopyTesting2004,
  title = {Black Hole Spectroscopy: {{Testing}} General Relativity through Gravitational Wave Observations},
  shorttitle = {Black-Hole Spectroscopy},
  author = {Dreyer, Olaf and Kelly, Bernard J. and Krishnan, Badri and Finn, Lee Samuel and Garrison, David and {Lopez-Aleman}, Ramon},
  year = 2004,
  journal = {Classical and Quantum Gravity},
  volume = {21},
  number = {4},
  eprint = {gr-qc/0309007},
  pages = {787--804},
  issn = {0264-9381},
  doi = {10.1088/0264-9381/21/4/003},
  urldate = {2025-02-07},
  archiveprefix = {arXiv},
  langid = {english}
}

@article{endlichEffectiveFormalismTesting2017,
  title = {An Effective Formalism for Testing Extensions to {{General Relativity}} with Gravitational Waves},
  author = {Endlich, Solomon and Gorbenko, Victor and Huang, Junwu and Senatore, Leonardo},
  year = 2017,
  month = sep,
  journal = {Journal of High Energy Physics},
  volume = {09},
  number = {9},
  eprint = {1704.01590},
  primaryclass = {gr-qc},
  pages = {122},
  issn = {1029-8479},
  doi = {10.1007/JHEP09(2017)122},
  urldate = {2025-02-06},
  archiveprefix = {arXiv},
  langid = {english}
}

@misc{fernandesLeadingEffectiveField2025,
  title = {Leading Effective Field Theory Corrections to the {{Kerr}} Metric at All Spins},
  author = {Fernandes, Pedro G.S.},
  year = 2025,
  month = dec,
  eprint = {2512.02338},
  primaryclass = {gr-qc},
  publisher = {arXiv},
  doi = {10.48550/arXiv.2512.02338},
  urldate = {2026-02-09},
  archiveprefix = {arXiv}
}

@incollection{franchiniTestingGeneralRelativity2024,
  title = {Testing {{General Relativity}} with {{Black Hole Quasi-normal Modes}}},
  booktitle = {Recent Progress on Gravity Tests: {{Challenges}} and Future Perspectives},
  author = {Franchini, Nicola and V{\"o}lkel, Sebastian H.},
  year = 2024,
  eprint = {2305.01696},
  primaryclass = {gr-qc},
  pages = {361--416},
  publisher = {Springer Nature Singapore},
  address = {Singapore},
  doi = {10.1007/978-981-97-2871-8\_9},
  archiveprefix = {arXiv},
  isbn = {978-981-97-2871-8}
}

@article{fundimplications,
  title = {Fundamental {{Physics Implications}} for {{Higher-Curvature Theories}} from {{Binary Black Hole Signals}} in the {{LIGO-Virgo Catalog GWTC-1}}},
  author = {Nair, Remya and Perkins, Scott and Silva, Hector O. and Yunes, Nicol{\'a}s},
  year = 2019,
  month = nov,
  journal = {Physical Review Letters},
  volume = {123},
  number = {19},
  eprint = {1905.00870},
  primaryclass = {gr-qc},
  pages = {191101},
  publisher = {American Physical Society},
  doi = {10.1103/PhysRevLett.123.191101},
  archiveprefix = {arXiv}
}

@article{grandclementSpectralMethodsNumerical2009,
  title = {Spectral Methods for Numerical Relativity},
  author = {Grandcl{\'e}ment, Philippe and Novak, J{\'e}r{\^o}me},
  year = 2009,
  journal = {Living Reviews in Relativity},
  volume = {12},
  number = {1},
  eprint = {0706.2286},
  primaryclass = {gr-qc},
  pages = {1},
  issn = {1433-8351},
  doi = {10.12942/lrr-2009-1},
  urldate = {2025-01-28},
  archiveprefix = {arXiv},
  langid = {english}
}

@article{greenAnomalyCancellationsSupersymmetric1984,
  title = {Anomaly {{Cancellation}} in {{Supersymmetric D}}=10 {{Gauge Theory}} and {{Superstring Theory}}},
  author = {Green, Michael B. and Schwarz, John H.},
  year = 1984,
  journal = {Physics Letters B},
  volume = {149},
  number = {1},
  pages = {117--122},
  issn = {0370-2693},
  doi = {10.1016/0370-2693(84)91565-X},
  urldate = {2026-03-10}
}

@article{horowitzamplification,
  title = {Extremal {{Kerr Black Holes}} as {{Amplifiers}} of {{New Physics}}},
  author = {Horowitz, Gary T. and Kolanowski, Maciej and Remmen, Grant N. and Santos, Jorge E.},
  year = 2023,
  month = aug,
  journal = {Physical Review Letters},
  volume = {131},
  number = {9},
  eprint = {2303.07358},
  primaryclass = {hep-th},
  pages = {091402},
  publisher = {American Physical Society},
  doi = {10.1103/PhysRevLett.131.091402},
  archiveprefix = {arXiv}
}

@article{hussainApproachComputingSpectral2022,
  title = {Approach to Computing Spectral Shifts for Black Holes beyond {{Kerr}}},
  author = {Hussain, Asad and Zimmerman, Aaron},
  year = 2022,
  month = nov,
  journal = {Physical Review D},
  volume = {106},
  number = {10},
  eprint = {2206.10653},
  primaryclass = {gr-qc},
  pages = {104018},
  issn = {2470-0010, 2470-0029},
  doi = {10.1103/PhysRevD.106.104018},
  urldate = {2026-03-20},
  archiveprefix = {arXiv}
}

@article{julieInspiralmergerringdownWaveforms2025,
  title = {Inspiral-Merger-Ringdown Waveforms in {{Einstein-scalar-Gauss-Bonnet}} Gravity within the Effective-One-Body Formalism},
  author = {Juli{\'e}, F{\'e}lix-Louis and Pompili, Lorenzo and Buonanno, Alessandra},
  year = 2025,
  month = jan,
  journal = {Physical Review D: Particles and Fields},
  volume = {111},
  number = {2},
  eprint = {2406.13654},
  primaryclass = {gr-qc},
  pages = {024016},
  doi = {10.1103/PhysRevD.111.024016},
  archiveprefix = {arXiv}
}

@article{konoplyaQuasinormalModesStability2020,
  title = {Quasinormal Modes, Stability and Shadows of a Black Hole in the {{4D Einstein}}--{{Gauss}}--{{Bonnet}} Gravity},
  author = {Konoplya, R.A. and Zinhailo, A.F.},
  year = 2020,
  month = nov,
  journal = {The European Physical Journal C: Particles and Fields},
  volume = {80},
  number = {11},
  eprint = {2003.01188},
  primaryclass = {gr-qc},
  pages = {1049},
  publisher = {American Physical Society},
  doi = {10.1140/epjc/s10052-020-08639-8},
  archiveprefix = {arXiv}
}

@article{lamAnalyticAccurateApproximate2026,
  title = {Analytic and Accurate Approximate Metrics for Black Holes with Arbitrary Rotation in Beyond-{{Einstein}} Gravity Using Spectral Methods},
  author = {Lam, Kelvin Ka-Ho and Chung, Adrian Ka-Wai and Yunes, Nicol{\'a}s},
  year = 2026,
  month = jan,
  journal = {Physical Review D},
  volume = {113},
  number = {2},
  eprint = {2510.05208},
  primaryclass = {gr-qc},
  pages = {024030},
  publisher = {American Physical Society},
  doi = {10.1103/txxb-z8bx},
  urldate = {2026-02-03},
  archiveprefix = {arXiv}
}

@article{lamSpinningBlackHoles2026,
  title = {Spinning {{Black Holes}} in {{Modified Gravity}} via {{Spectral Methods}}},
  author = {Lam, Kelvin Ka-Ho and Chung, Adrian Ka-Wai and Yunes, Nicol{\'a}s},
  year = 2026,
  month = jan,
  journal = {Physical Review Letters},
  volume = {136},
  number = {2},
  eprint = {2509.07061},
  primaryclass = {gr-qc},
  pages = {021401},
  publisher = {American Physical Society},
  doi = {10.1103/6z43-xdv7},
  urldate = {2026-02-03},
  archiveprefix = {arXiv}
}

@article{leaverAnalyticRepresentationQuasinormal1985,
  title = {An {{Analytic}} Representation for the Quasi Normal Modes of {{Kerr}} Black Holes},
  author = {Leaver, E.W.},
  year = 1985,
  journal = {Proceedings of the Royal Society of London. A. Mathematical and Physical Sciences},
  volume = {402},
  number = {1823},
  pages = {285--298},
  issn = {0080-4630},
  doi = {10.1098/rspa.1985.0119},
  urldate = {2025-12-11}
}

@article{liPerturbationsSpinningBlack2023,
  title = {Perturbations of {{Spinning Black Holes}} beyond {{General Relativity}}: {{Modified Teukolsky Equation}}},
  shorttitle = {Perturbations of Spinning Black Holes beyond {{General Relativity}}},
  author = {Li, Dongjun and Wagle, Pratik and Chen, Yanbei and Yunes, Nicol{\'a}s},
  year = 2023,
  month = apr,
  journal = {Physical Review X},
  volume = {13},
  number = {2},
  eprint = {2206.10652},
  primaryclass = {gr-qc},
  pages = {021029},
  issn = {2160-3308},
  doi = {10.1103/PhysRevX.13.021029},
  urldate = {2026-03-20},
  archiveprefix = {arXiv}
}

@article{liuRobustImprovedConstraints2025,
  title = {Robust and Improved Constraints on Higher-Curvature Gravitational Effective-Field-Theory with the {{GW170608}} Event},
  author = {Liu, Haoyang and Yunes, Nicol{\'a}s},
  year = 2025,
  month = apr,
  journal = {Physical Review D: Particles and Fields},
  volume = {111},
  number = {8},
  eprint = {2407.08929},
  primaryclass = {gr-qc},
  pages = {084049},
  doi = {10.1103/PhysRevD.111.084049},
  archiveprefix = {arXiv}
}

@article{loustoRemnantMassesSpins2010,
  title = {Remnant {{Masses}}, {{Spins}} and {{Recoils}} from the {{Merger}} of {{Generic Black-Hole Binaries}}},
  author = {Lousto, Carlos O. and Campanelli, Manuela and Zlochower, Yosef and Nakano, Hiroyuki},
  year = 2010,
  journal = {Classical and Quantum Gravity},
  volume = {27},
  number = {11},
  eprint = {0904.3541},
  primaryclass = {gr-qc},
  pages = {114006},
  issn = {0264-9381, 1361-6382},
  doi = {10.1088/0264-9381/27/11/114006},
  urldate = {2026-03-11},
  archiveprefix = {arXiv}
}

@article{LVKcollaborationBlackHoleSpectroscopy2026,
  title = {Black {{Hole Spectroscopy}} and {{Tests}} of {{General Relativity}} with {{GW250114}}},
  author = {Abac, A.G. and Abouelfettouh, I. and Acernese, F. and others},
  year = 2026,
  month = jan,
  journal = {Physical Review Letters},
  volume = {136},
  number = {4},
  eprint = {2509.08099},
  primaryclass = {gr-qc},
  pages = {041403},
  publisher = {American Physical Society},
  doi = {10.1103/6c61-fm1n},
  urldate = {2026-03-09},
  archiveprefix = {arXiv},
  collaboration = {LIGO-Virgo-KAGRA Collaboration}
}

@article{LVKcollaborationGW250114TestingHawkings2025,
  title = {{{GW250114}}: {{Testing Hawking}}'s {{Area Law}} and the {{Kerr Nature}} of {{Black Holes}}},
  shorttitle = {{{GW250114}}},
  author = {Abac, A.G. and Abouelfettouh, I. and Acernese, F. and others},
  year = 2025,
  month = sep,
  journal = {Physical Review Letters},
  volume = {135},
  number = {11},
  eprint = {2509.08054},
  primaryclass = {gr-qc},
  pages = {111403},
  publisher = {American Physical Society},
  doi = {10.1103/kw5g-d732},
  urldate = {2025-09-12},
  archiveprefix = {arXiv},
  collaboration = {LIGO-Virgo-KAGRA Collaboration}
}

@misc{LVKcollaborationGWTC40UpdatingGravitationalWave2025,
  title = {{{GWTC-4}}.0: {{Updating}} the {{Gravitational-Wave Transient Catalog}} with {{Observations}} from the {{First Part}} of the {{Fourth LIGO-Virgo-KAGRA Observing Run}}},
  shorttitle = {{{GWTC-4}}.0},
  author = {Abac, A.G. and Abouelfettouh, I. and Acernese, F. and others},
  year = 2025,
  month = aug,
  eprint = {2508.18082},
  primaryclass = {gr-qc},
  publisher = {arXiv},
  doi = {10.48550/arXiv.2508.18082},
  urldate = {2026-03-09},
  archiveprefix = {arXiv},
  collaboration = {LIGO-Virgo-KAGRA Collaboration}
}

@misc{LVKcollaborationTestsGeneralRelativity2021,
  title = {Tests of General Relativity with Binary Black Holes from the Second {{LIGO-Virgo}} Gravitational-Wave Transient Catalog},
  author = {Abbott, R. and others},
  year = 2021,
  volume = {103},
  number = {LIGO-P2000091},
  eprint = {2010.14529},
  primaryclass = {gr-qc},
  pages = {122002},
  doi = {10.1103/PhysRevD.103.122002},
  archiveprefix = {arXiv},
  collaboration = {LIGO Scientific, Virgo}
}

@misc{LVKcollaborationTestsGeneralRelativity2025,
  title = {Tests of {{General Relativity}} with {{GWTC-3}}},
  author = {Abbott, R. and Abe, H. and Acernese, F. and others},
  year = 2025,
  month = oct,
  volume = {112},
  number = {8},
  pages = {084080},
  issn = {2470-0010, 2470-0029},
  doi = {10.1103/PhysRevD.112.084080},
  collaboration = {LIGO-Virgo-KAGRA Collaboration}
}

@misc{LVKcollaborationTestsGeneralRelativity2026a,
  title = {{{GWTC-4}}.0: {{Tests}} of {{General Relativity}}. {{I}}. {{Overview}} and {{General Tests}}},
  author = {Abac, A.G. and Abouelfettouh, I. and Acernese, F. and others},
  year = 2026,
  month = mar,
  eprint = {2603.19019},
  primaryclass = {gr-qc},
  archiveprefix = {arXiv},
  collaboration = {LIGO-Virgo-KAGRA Collaboration}
}

@misc{LVKcollaborationTestsGeneralRelativity2026b,
  title = {{{GWTC-4}}.0: {{Tests}} of {{General Relativity}}. {{II}}. {{Parameterized Tests}}},
  author = {Abac, A.G. and Abouelfettouh, I. and Acernese, F. and others},
  year = 2026,
  month = mar,
  eprint = {2603.19020},
  primaryclass = {gr-qc},
  archiveprefix = {arXiv},
  collaboration = {LIGO-Virgo-KAGRA Collaboration}
}

@misc{LVKcollaborationTestsGeneralRelativity2026c,
  title = {{{GWTC-4}}.0: {{Tests}} of {{General Relativity}}. {{III}}. {{Tests}} of the {{Remnants}}},
  author = {Abac, A.G. and Abouelfettouh, I. and Acernese, F. and others},
  year = 2026,
  month = mar,
  eprint = {2603.19021},
  primaryclass = {gr-qc},
  archiveprefix = {arXiv},
  collaboration = {LIGO Scientific, VIRGO, KAGRA}
}

@article{maedaBlackHoleSolutions2009,
  title = {Black {{Hole Solutions}} in {{String Theory}} with {{Gauss-Bonnet Curvature Correction}}},
  author = {Maeda, Kei-ichi and Ohta, Nobuyoshi and Sasagawa, Yukinori},
  year = 2009,
  journal = {Physical Review D},
  volume = {80},
  number = {10},
  eprint = {0908.4151},
  primaryclass = {hep-th},
  pages = {104032},
  issn = {1550-7998, 1550-2368},
  doi = {10.1103/PhysRevD.80.104032},
  urldate = {2026-03-10},
  archiveprefix = {arXiv}
}

@article{maenautRingdownAnalysisRotating2026,
  title = {Ringdown Analysis of Rotating Black Holes in Effective Field Theory Extensions of General Relativity},
  author = {Maenaut, Simon and Carullo, Gregorio and Cano, Pablo A. and Liu, Anna and Cardoso, Vitor and Hertog, Thomas and Li, Tjonnie G.F.},
  year = 2026,
  month = feb,
  journal = {Physical Review D: Particles and Fields},
  volume = {113},
  number = {4},
  eprint = {2411.17893},
  primaryclass = {gr-qc},
  pages = {044039},
  doi = {10.1103/k3hv-cqfn},
  archiveprefix = {arXiv}
}

@article{maselliRotatingBlackHoles2015,
  title = {Rotating Black Holes in {{Einstein-Dilaton-Gauss-Bonnet}} Gravity with Finite Coupling},
  author = {Maselli, Andrea and Pani, Paolo and Gualtieri, Leonardo and Ferrari, Valeria},
  year = 2015,
  month = oct,
  journal = {Physical Review D},
  volume = {92},
  number = {8},
  eprint = {1507.00680},
  primaryclass = {gr-qc},
  pages = {083014},
  publisher = {American Physical Society},
  doi = {10.1103/PhysRevD.92.083014},
  urldate = {2026-03-11},
  archiveprefix = {arXiv}
}

@article{mcmanusParametrizedBlackholeQuasinormal2019,
  title = {Parametrized Black Hole Quasinormal Ringdown. {{II}}. {{Coupled}} Equations and Quadratic Corrections for Nonrotating Black Holes},
  author = {McManus, Ryan and Berti, Emanuele and Macedo, Caio F.B. and Kimura, Masashi and Maselli, Andrea and Cardoso, Vitor},
  year = 2019,
  month = aug,
  journal = {Physical Review D: Particles and Fields},
  volume = {100},
  number = {4},
  eprint = {1906.05155},
  primaryclass = {gr-qc},
  pages = {044061},
  publisher = {American Physical Society},
  doi = {10.1103/PhysRevD.100.044061},
  archiveprefix = {arXiv}
}

@article{miguelEFTCorrectionsScalar2024,
  title = {{{EFT}} Corrections to Scalar and Vector Quasinormal Modes of Rapidly Rotating Black Holes},
  author = {Miguel, Filipe S.},
  year = 2024,
  month = may,
  journal = {Physical Review D},
  volume = {109},
  number = {10},
  eprint = {2308.03832},
  primaryclass = {gr-qc},
  pages = {104016},
  publisher = {American Physical Society},
  doi = {10.1103/PhysRevD.109.104016},
  urldate = {2025-11-30},
  archiveprefix = {arXiv}
}

@article{molinaGravitationalSignatureSchwarzschild2010,
  title = {Gravitational Signature of {{Schwarzschild}} Black Holes in Dynamical {{Chern-Simons}} Gravity},
  author = {Molina, C. and Pani, Paolo and Cardoso, Vitor and Gualtieri, Leonardo},
  year = 2010,
  journal = {Physical Review D: Particles and Fields},
  volume = {81},
  eprint = {1004.4007},
  primaryclass = {gr-qc},
  pages = {124021},
  doi = {10.1103/PhysRevD.81.124021},
  archiveprefix = {arXiv}
}

@article{mouraHigherDerivativeCorrectedBlack2007,
  title = {Higher-Derivative Corrected Black Holes: {{Perturbative}} Stability and Absorption Cross-Section in Heterotic String Theory},
  shorttitle = {Higher-{{Derivative Corrected Black Holes}}},
  author = {Moura, Filipe and Schiappa, Ricardo},
  year = 2007,
  journal = {Classical and Quantum Gravity},
  volume = {24},
  number = {2},
  eprint = {hep-th/0605001},
  pages = {361--386},
  issn = {0264-9381, 1361-6382},
  doi = {10.1088/0264-9381/24/2/006},
  urldate = {2026-03-10},
  archiveprefix = {arXiv}
}

@article{no_hair,
  title = {Event Horizons in Static Vacuum Space-Times},
  author = {Israel, Werner},
  year = 1967,
  journal = {Physical Review},
  volume = {164},
  number = {5},
  pages = {1776--1779},
  publisher = {American Physical Society},
  doi = {10.1103/PhysRev.164.1776}
}

@article{no_hair2,
  title = {Event Horizons in Static Electrovac Space-Times},
  author = {Israel, Werner},
  year = 1968,
  journal = {Communications in Mathematical Physics},
  volume = {8},
  number = {3},
  pages = {245--260},
  doi = {10.1007/BF01645859}
}

@article{no_hair3,
  title = {Axisymmetric {{Black Hole Has Only Two Degrees}} of {{Freedom}}},
  author = {Carter, B.},
  year = 1971,
  journal = {Physical Review Letters},
  volume = {26},
  number = {6},
  pages = {331--333},
  publisher = {American Physical Society},
  doi = {10.1103/PhysRevLett.26.331}
}

@misc{paniScalarShortcutBeyondKerr2026,
  title = {Scalar Shortcut to Beyond-{{Kerr}} Ringdown Tests and Their Complementarity with Black-Hole Shadow Observations},
  author = {Pani, Paolo and Sanna, Andrea P.},
  year = 2026,
  month = mar,
  eprint = {2603.08782},
  primaryclass = {gr-qc},
  publisher = {arXiv},
  doi = {10.48550/arXiv.2603.08782},
  urldate = {2026-03-13},
  archiveprefix = {arXiv}
}

@article{pieriniQuasinormalmodesRotatingBlackholes2021,
  title = {Quasi-Normal Modes of Rotating Black Holes in {{Einstein-dilaton Gauss-Bonnet}} Gravity: The First Order in Rotation},
  author = {Pierini, Lorenzo and Gualtieri, Leonardo},
  year = 2021,
  month = jun,
  journal = {Physical Review D: Particles and Fields},
  volume = {103},
  number = {12},
  eprint = {2103.09870},
  primaryclass = {gr-qc},
  pages = {124017},
  publisher = {American Physical Society},
  doi = {10.1103/PhysRevD.103.124017},
  archiveprefix = {arXiv}
}

@article{pieriniQuasinormalmodesRotatingBlackholes2022,
  title = {Quasinormal Modes of Rotating Black Holes in {{Einstein-dilaton Gauss-Bonnet}} Gravity: {{The}} Second Order in Rotation},
  author = {Pierini, Lorenzo and Gualtieri, Leonardo},
  year = 2022,
  month = nov,
  journal = {Physical Review D: Particles and Fields},
  volume = {106},
  number = {10},
  eprint = {2207.11267},
  primaryclass = {gr-qc},
  pages = {104009},
  publisher = {American Physical Society},
  doi = {10.1103/PhysRevD.106.104009},
  archiveprefix = {arXiv}
}

@article{pratikQuasinormalModesSlowlyrotating2022,
  title = {Quasinormal Modes of Slowly-Rotating Black Holes in Dynamical {{Chern-Simons}} Gravity},
  author = {Wagle, Pratik and Yunes, Nicol{\'a}s and Silva, Hector O.},
  year = 2022,
  month = jun,
  journal = {Physical Review D: Particles and Fields},
  volume = {105},
  number = {12},
  eprint = {2103.09913},
  primaryclass = {gr-qc},
  pages = {124003},
  publisher = {American Physical Society},
  doi = {10.1103/PhysRevD.105.124003},
  archiveprefix = {arXiv}
}

@article{pressPerturbationsRotatingBlack1973,
  title = {Perturbations of a {{Rotating Black Hole}}. {{II}}. {{Dynamical Stability}} of the {{Kerr Metric}}},
  author = {Press, William H. and Teukolsky, Saul A.},
  year = 1973,
  journal = {The Astrophysical Journal},
  volume = {185},
  pages = {649--674},
  publisher = {IOP},
  issn = {0004-637X},
  doi = {10.1086/152445},
  urldate = {2024-09-11}
}

@misc{sangerTestsGeneralRelativity2024,
  title = {Tests of {{General Relativity}} with {{GW230529}}: A Neutron Star Merging with a Lower Mass-Gap Compact Object},
  shorttitle = {Tests of {{General Relativity}} with {{GW230529}}},
  author = {S{\"a}nger, Elise M. and Roy, Soumen and Agathos, Michalis and others},
  year = 2024,
  month = jun,
  eprint = {2406.03568},
  primaryclass = {gr-qc},
  publisher = {arXiv},
  doi = {10.48550/arXiv.2406.03568},
  urldate = {2026-03-25},
  archiveprefix = {arXiv}
}

@article{silvaAstrophysicalTheoreticalPhysics2021,
  title = {Astrophysical and Theoretical Physics Implications from Multimessenger Neutron Star Observations},
  author = {Silva, Hector O. and Holgado, A. Miguel and {C{\'a}rdenas-Avenda{\~n}o}, Alejandro and Yunes, Nicol{\'a}s},
  year = 2021,
  month = may,
  journal = {Physical Review Letters},
  volume = {126},
  number = {18},
  eprint = {2004.01253},
  primaryclass = {gr-qc},
  pages = {181101},
  issn = {0031-9007, 1079-7114},
  doi = {10.1103/PhysRevLett.126.181101},
  urldate = {2026-03-31},
  archiveprefix = {arXiv}
}

@article{silvaBlackholeRingdownProbe2023,
  title = {Black-Hole Ringdown as a Probe of Higher-Curvature Gravity Theories},
  author = {Silva, Hector O. and Ghosh, Abhirup and Buonanno, Alessandra},
  year = 2023,
  month = feb,
  journal = {Physical Review D: Particles and Fields},
  volume = {107},
  number = {4},
  eprint = {2205.05132},
  primaryclass = {gr-qc},
  pages = {044030},
  doi = {10.1103/PhysRevD.107.044030},
  archiveprefix = {arXiv}
}

@article{silvaQuasinormalModesExcitation2024,
  title = {Quasinormal Modes and Their Excitation beyond General Relativity},
  author = {Silva, Hector O. and Tambalo, Giovanni and Glampedakis, Kostas and Yagi, Kent and Steinhoff, Jan},
  year = 2024,
  month = jul,
  journal = {Physical Review D: Particles and Fields},
  volume = {110},
  number = {2},
  eprint = {2404.11110},
  primaryclass = {gr-qc},
  pages = {024042},
  publisher = {American Physical Society},
  doi = {10.1103/PhysRevD.110.024042},
  archiveprefix = {arXiv}
}

@article{srivastavaAnalyticalComputationQuasinormalmodes2021,
  title = {Analytical Computation of Quasinormal Modes of Slowly Rotating Black Holes in Dynamical {{Chern-Simons}} Gravity},
  author = {Srivastava, Manu and Chen, Yanbei and Shankaranarayanan, S.},
  year = 2021,
  month = sep,
  journal = {Physical Review D: Particles and Fields},
  volume = {104},
  number = {6},
  eprint = {2106.06209},
  primaryclass = {gr-qc},
  pages = {064034},
  publisher = {American Physical Society},
  doi = {10.1103/PhysRevD.104.064034},
  archiveprefix = {arXiv}
}

@article{stelleClassicalGravityHigher1978,
  title = {Classical {{Gravity}} with {{Higher Derivatives}}},
  author = {Stelle, K. S.},
  year = 1978,
  journal = {General Relativity and Gravitation},
  volume = {9},
  number = {4},
  pages = {353--371},
  issn = {1572-9532},
  doi = {10.1007/BF00760427},
  urldate = {2026-03-02},
  langid = {english}
}

@article{stelleRenormalizationHigherderivativeQuantum1977,
  title = {Renormalization of {{Higher Derivative Quantum Gravity}}},
  author = {Stelle, K.S.},
  year = 1977,
  journal = {Physical Review D},
  volume = {16},
  number = {4},
  pages = {953--969},
  publisher = {American Physical Society},
  doi = {10.1103/PhysRevD.16.953},
  urldate = {2026-03-10}
}

@article{tattersallQuasinormalmodesBlackholesHorndeski2018,
  title = {Quasinormal Modes of Black Holes in {{Horndeski}} Gravity},
  author = {Tattersall, Oliver J. and Ferreira, Pedro G.},
  year = 2018,
  month = may,
  journal = {Physical Review D: Particles and Fields},
  volume = {97},
  number = {10},
  eprint = {1804.08950},
  primaryclass = {gr-qc},
  pages = {104047},
  publisher = {American Physical Society},
  doi = {10.1103/PhysRevD.97.104047},
  archiveprefix = {arXiv}
}

@article{teukolskyPerturbationsRotatingBlack1973,
  title = {Perturbations of a Rotating Black Hole. 1. {{Fundamental}} Equations for Gravitational Electromagnetic and {$\nu$} Field Perturbations},
  author = {Teukolsky, Saul A.},
  year = 1973,
  month = oct,
  journal = {The Astrophysical Journal},
  volume = {185},
  pages = {635--647},
  publisher = {IOP},
  issn = {0004-637X},
  doi = {10.1086/152444},
  urldate = {2024-09-11}
}

@article{teukolskyRotatingBlackHoles1972,
  title = {Rotating Black Holes - Separable Wave Equations for Gravitational and Electromagnetic Perturbations},
  shorttitle = {Rotating {{Black Holes}}},
  author = {Teukolsky, S.A.},
  year = 1972,
  journal = {Physical Review Letters},
  volume = {29},
  number = {16},
  pages = {1114--1118},
  publisher = {American Physical Society},
  doi = {10.1103/PhysRevLett.29.1114},
  urldate = {2024-07-11}
}

@article{vishveshwaraStabilitySchwarzschildMetric1970,
  title = {Stability of the Schwarzschild Metric},
  author = {Vishveshwara, C.V.},
  year = 1970,
  journal = {Physical Review D},
  volume = {1},
  number = {10},
  pages = {2870--2879},
  publisher = {American Physical Society},
  doi = {10.1103/PhysRevD.1.2870},
  urldate = {2024-09-11}
}

@article{waglePerturbationsSpinningBlack2024,
  title = {Perturbations of Spinning Black Holes in Dynamical {{Chern-Simons}} Gravity: {{Slow}} Rotation Equations},
  shorttitle = {Perturbations of Spinning Black Holes in Dynamical {{Chern-Simons}} Gravity},
  author = {Wagle, Pratik and Li, Dongjun and Chen, Yanbei and Yunes, Nicol{\'a}s},
  year = 2024,
  month = may,
  journal = {Physical Review D},
  volume = {109},
  number = {10},
  eprint = {2311.07706},
  primaryclass = {gr-qc},
  pages = {104029},
  publisher = {American Physical Society},
  doi = {10.1103/PhysRevD.109.104029},
  urldate = {2026-03-20},
  archiveprefix = {arXiv}
}

@article{yagiSlowlyRotatingBlack2012,
  title = {Slowly Rotating Black Holes in Dynamical {{Chern-Simons}} Gravity: {{Deformation}} Quadratic in the Spin},
  shorttitle = {Slowly Rotating Black Holes in Dynamical {{Chern-Simons}} Gravity},
  author = {Yagi, Kent and Yunes, Nicol{\'a}s and Tanaka, Takahiro},
  year = 2012,
  month = aug,
  journal = {Physical Review D},
  volume = {86},
  number = {4},
  eprint = {1206.6130},
  primaryclass = {gr-qc},
  pages = {044037},
  publisher = {American Physical Society},
  doi = {10.1103/PhysRevD.86.044037},
  urldate = {2026-03-11},
  archiveprefix = {arXiv}
}

@article{yangBranchingQuasinormalModes2013,
  title = {Branching of Quasinormal Modes for Nearly Extremal {{Kerr}} Black Holes},
  author = {Yang, Huan and Zhang, Fan and Zimmerman, Aaron and Nichols, David A. and Berti, Emanuele and Chen, Yanbei},
  year = 2013,
  month = feb,
  journal = {Physical Review D},
  volume = {87},
  number = {4},
  eprint = {1212.3271},
  primaryclass = {gr-qc},
  pages = {041502},
  publisher = {American Physical Society},
  doi = {10.1103/PhysRevD.87.041502},
  urldate = {2026-01-29},
  archiveprefix = {arXiv}
}

@article{yangQuasinormalModesNearly2013,
  title = {Quasinormal Modes of Nearly Extremal {{Kerr}} Spacetimes: Spectrum Bifurcation and Power-Law Ringdown},
  shorttitle = {Quasinormal Modes of Nearly Extremal {{Kerr}} Spacetimes},
  author = {Yang, Huan and Zimmerman, Aaron and Zengino{\u g}lu, An{\i}l and Zhang, Fan and Berti, Emanuele and Chen, Yanbei},
  year = 2013,
  month = aug,
  journal = {Physical Review D},
  volume = {88},
  number = {4},
  eprint = {1307.8086},
  primaryclass = {gr-qc},
  pages = {044047},
  publisher = {American Physical Society},
  doi = {10.1103/PhysRevD.88.044047},
  urldate = {2026-01-29},
  archiveprefix = {arXiv}
}

@article{yunesDynamicalChernSimonsModified2009,
  title = {Dynamical {{Chern-Simons Modified Gravity}}. {{I}}. {{Spinning Black Holes}} in the {{Slow-Rotation Approximation}}},
  shorttitle = {Dynamical {{Chern-Simons}} Modified Gravity},
  author = {Yunes, Nicol{\'a}s and Pretorius, Frans},
  year = 2009,
  journal = {Physical Review D},
  volume = {79},
  number = {8},
  eprint = {0902.4669},
  primaryclass = {gr-qc},
  pages = {084043},
  publisher = {American Physical Society},
  doi = {10.1103/PhysRevD.79.084043},
  urldate = {2026-03-11},
  archiveprefix = {arXiv}
}

@article{yunesGravitationalwaveTestsGeneral2025,
  title = {Gravitational-Wave Tests of General Relativity with Ground-Based Detectors and Pulsar-Timing Arrays},
  author = {Yunes, Nicol{\'a}s and Siemens, Xavier and Yagi, Kent},
  year = 2025,
  month = mar,
  journal = {Living Reviews in Relativity},
  volume = {28},
  number = {1},
  pages = {3},
  issn = {1433-8351},
  doi = {10.1007/s41114-024-00054-9},
  urldate = {2026-03-09},
  langid = {english}
}

@misc{zajacek2019electricchargeblackholes,
  title = {Electric Charge of Black Holes: {{Is}} It Really Always Negligible?},
  author = {Zaja{\v c}ek, Michal and Tursunov, Arman},
  year = 2019,
  month = apr,
  eprint = {1904.04654},
  primaryclass = {astro-ph.GA},
  archiveprefix = {arXiv}
}

@inproceedings{zimmermanQuasinormalModesKerr2015,
  title = {Quasinormal {{Modes Beyond Kerr}}},
  booktitle = {Gravitational {{Wave Astrophysics}}},
  author = {Zimmerman, Aaron and Yang, Huan and Mark, Zachary and Chen, Yanbei and Lehner, Luis},
  editor = {Sopuerta, Carlos F.},
  year = 2015,
  pages = {217--223},
  publisher = {Springer International Publishing},
  address = {Cham},
  doi = {10.1007/978-3-319-10488-1\_19},
  urldate = {2026-03-06},
  isbn = {978-3-319-10488-1},
  langid = {english}
}

% \clearpage

{\onecolumngrid
\appendix

\section{Tabulated scalar QNMs}

We show the values of the scalar QNMs and their lowest order corrections in quadratic gravity in Table \ref{tab:qnms}. One can see the amplification of certain modes in the large spin limit. Parentheses denote uncertainties in the numerical results.

\begin{table*}[h]
\centering
\begin{ruledtabular}
\begin{tabular}{ccc|d|dd|dd|dd}
   $l$ & $m$  &  $n$ &  \multicolumn{1}{c|}{$a$} 
   & \multicolumn{1}{r}{$\Re[\omega^{(0)}]$} 
   & \multicolumn{1}{r|}{$\Im[\omega^{(0)}]$}
   & \multicolumn{1}{r}{$\Re[\omega^{(1)}_\mathrm{sGB}]$} 
   & \multicolumn{1}{r|}{$\Im[\omega^{(1)}_\mathrm{sGB}]$}
   & \multicolumn{1}{r}{$\Re[\omega^{(1)}_\mathrm{dCS}]$} 
   & \multicolumn{1}{r}{$\Im[\omega^{(1)}_\mathrm{dCS}]$}\\
   \hline
   \hline
   \multirow{5}{*}{$0$} & \multirow{5}{*}{$0$} & \multirow{5}{*}{$0$} & 
   0 & 0.110455 & -0.104896 & 0.051470 & 0.001310 & 0 & 0\\ 
   & & & 0.5  & 0.112381 & -0.102183 & 0.05139(1) & 0.00849(6) & 0.009567 & 0.000840\\
   & & & 0.7  & 0.113979 & -0.098632 & 0.04991(9) & 0.01646(7) & 0.02488(2) & 0.00511(0)\\
   & & & 0.9  & 0.113848 & -0.091569 & 0.0576(21) & 0.0284(34) & 0.0766(41) & 0.0422(69)\\
   & & & 0.99 & 0.110447 & -0.089499 & -0.0014(33) & 0.0719(87)  & -0.00144(7) & 0.14071(8)\\
   \hline
   \multirow{5}{*}{$1$} & \multirow{5}{*}{$1$} & \multirow{5}{*}{$0$} & 
   0 & 0.292936 & -0.097660 & 0.070926 & 0.006985 & 0 & 0\\ 
   & & & 0.5  & 0.344753 & -0.094395 & 0.121804 & 0.007025 & -0.020141 & -0.032287\\
   & & & 0.7  & 0.379159 & -0.088848 & 0.142872 & -0.003041 & -0.035884 & -0.067110\\
   & & & 0.9  & 0.437234 & -0.071848 & 0.056850 & -0.149358 & -0.115447 & -0.233875\\
   & & & 0.99 & 0.493423 & -0.036712 & -1.128086 & -2.533662  & -0.761281 & -1.976180\\
   \hline
   \multirow{5}{*}{$2$} & \multirow{5}{*}{$2$} & \multirow{5}{*}{$0$} & 
    0 & 0.483644 & -0.096759 & 0.104475 & 0.007850 & 0 & 0\\ 
   & & & 0.5  & 0.585990 & -0.093494 & 0.201456 & 0.007498 & -0.041427 & -0.042289\\
   & & & 0.7  & 0.656099 & -0.087649 & 0.237865 & -0.008203 & -0.087212 & -0.091236\\
   & & & 0.9  & 0.781638 & -0.069289 & -0.036615 & -0.194042 & -0.387408 & -0.308019\\
   & & & 0.99 & 0.928028 & -0.031063 & -5.511824 & -2.784240  & -4.110527 & -2.209703\\
   \hline
   \multirow{5}{*}{$2$} & \multirow{5}{*}{$2$} & \multirow{5}{*}{$1$} & 
    0 & 0.463851 & -0.295604 & 0.149540 & 0.018393 & 0 & 0\\ 
   & & & 0.5  & 0.573443 & -0.283336 & 0.254432 & 0.012974 & -0.051775 & -0.135720\\
   & & & 0.7  & 0.647452 & -0.264560 & 0.279373 & -0.037184 & -0.095981 & -0.290897\\
   & & & 0.9  & 0.777683 & -0.208009 & -0.069565 & 0.598631 & -0.403820 & -0.959286\\
   & & & 0.99 & 0.926686 & -0.092600 & -5.837081 & -8.240921 & -4.346625 & -6.555512\\
\end{tabular}
\end{ruledtabular}
\caption{Scalar QNM frequencies and the corresponding lowest order corrections of sGB and dCS for different modes and different spins. The table gives the real and imaginary part of the GR-frequency, and the sGB and dCS corrections for the $(l,m,n)=(0,0,0), (1,1,0),(2,2,0),(2,2,1)$ modes. Digits in parentheses indicate values beyond the confirmed significant digits, as determined from convergence analysis.}
\label{tab:qnms}
\end{table*}
}
\end{document}